\documentclass[a4paper,11pt]{article}
\usepackage{jinstpub} % for details on the use of the package, please see the JINST-author-manual
\usepackage{lineno}
\usepackage{multirow}

\usepackage{subcaption} 
\usepackage{caption} 
\title{\boldmath Classification of Hoyle State Decay Branches in Active Target Time Projection Chamber using Neural Network}

\author[a,b]{Pralay Kumar Das,}
\affiliation[a]{Saha Institute of Nuclear Physics, AF Block, Sector I, Bidhannagar, Kolkata 700064, India}
\affiliation[b]{Homi Bhabha National Institute, Training School Complex, Anushaktinagar, Mumbai 400094, India}
\affiliation[*]{Retired}
\emailAdd{pralay.das@saha.ac.in}
\author[a,b]{Nayana Majumdar}
\author[a,b,*]{and Supratik Mukhopadhyay}

\abstract{
A multi-class convolutional neural network model has been developed using the Keras deep learning library in Python for image-based classification of $^{12}$C Hoyle state decay branches from tracking information, to be recorded by a prototype Active Target Time Projection Chamber, SAT-TPC (currently under development), filled with an Ar + CO$_2$ (90:10) gas mixture at atmospheric pressure. The elastic scattering and Hoyle state sequential and direct decay events, in the interaction of 30 MeV $\alpha$-particle beam with $^{40}$Ar, $^{12}$C, $^{16}$O nuclei, have been generated through Monte-Carlo simulation. The three-dimensional tracks, produced by the scattering and decay products through primary ionization of the gaseous medium, have been simulated with Geant4. The primary tracks, distributed on the beam-plane, have been convoluted with electron diffusion, obtained with Magboltz, to produce the final tracking information. The classification performance of the proposed model for different readout segmentation schemes of the SAT-TPC has been discussed.}

\begin{document}
\maketitle
\flushbottom

\section{Introduction}
\label{sec:intro}
The formation of $^{12}$C by nuclear fusion of three $\alpha$-nuclei is a significant step in the stellar nucleo-synthesis. It plays an important role in producing the elements heavier than $^4$He and bridge the well-known gap, caused by the instability of nuclei of atomic weight 5 and 8. To explain the observed abundance of $^{12}$C in stellar media, the reaction $^{8}Be(\alpha,\gamma)^{12}C$ at the second step of the triple-$\alpha$ fusion reaction was proposed to proceed through {\it{s}}-wave resonance around 7.68 MeV above the ground level in $^{12}$C \cite{Hoyle1954}. An excited state with spin-parity $0^+$ and energy 7.65 MeV, namely the Hoyle state, was later verified experimentally \cite{Cook1957}.  Aside from its key role in the synthesis of the elements, there are several intriguing aspects about the structure of the Hoyle state. It has been theoretically conjectured as a molecule-like $\alpha$-cluster state \cite{Kami1981, Cher2007}, or Bose-Einstein Condensate (BEC)-like configuration \cite{Toh2001}, etc. Eventually, many experimental explorations as well as theoretical modelling have followed suit to envisage the exotic structure of the Hoyle state \cite{Free2014}. In this context, the decay modes of the Hoyle state serve as valuable observables. The relative branching ratio between the direct triple-$\alpha$ decay of $^{12}$C$^{\ast}$ and the sequential two-step decay via the ground state of $^8$Be can be theoretically estimated using a BEC-based wave function that includes two and three-body tunnelling processes \cite{Smith2017}. This investigation has an impact in understanding the stellar nucleosynthesis process as well. Any deviation from the sequential decay should modify the calculation of the nucleosynthesis reaction rate where exclusively the two-step sequential fusion is taken into account. Several experiments have been carried out to determine the branching ratio of the direct decay which has been found to be as small as 0.019\% by one of the latest measurements \cite{Rana2019}. The said limit has further been modified to 0.00057\% (5.7 $\times$ 10$^{-6}$)
following indirect measurements \cite{Smith2020}. These reports imply the necessity of measuring the cross-section of the Hoyle state decay branches to determine their branching ratio more precisely.

The Time Projection Chamber (TPC) \cite{Nyg1974, Mar1978}, equipped with a long drift volume and position-sensitive electron multiplier readout, is a stand-alone device that can provide three-dimensional (3D) position information of charged particles. It utilizes the drift time together with the two-dimensional (2D) position information of the electrons, produced in the primary ionization of the gaseous molecules of the active medium by the passage of the particle to produce the 3D position. In addition to wide usage in high-energy particle physics experiments, eventually it has found application in low-energy nuclear physics experiments as well for measuring nuclear scattering cross-sections. The relevant kinematics can be studied by 3D tracking of the scattering products using the same working principle of the TPC. However, the only change for these applications is implemented by opting the active gas medium as the target nucleus. Thus, the device, known as an Active Target Time Projection Chamber (AT-TPC) \cite{Ayyad2018}, can offer advantages to circumvent the issues related to the 
use of solid targets and elaborate detector arrangements. It may be further beneficial in evading the limitations of background, associated
with the use of silicon detector arrays in measuring very low branching ratios, as pertinent to the exploration of the Hoyle state configuration. A high-sensitivity measurement of the direct decay mode has been performed
recently using an AT-TPC \cite{Bis2020}. By applying a Bayesian approach to study the contribution of the direct decay using a likelihood
function, a limit of the direct decay mode less than 0.043\% at the 95\% confidence level has been achieved. 

%because they allow precise measurements of its decay pathways, including direct and sequential decays. These devices provide high sensitivity to low-energy particles and allow tracking \cite{Att2009} of reaction events in three dimensions, which is essential for disentangling the decay mechanisms. By combining a gas-filled detector with real-time tracking and energy measurements, active target TPCs enhance the ability to measure rare decay processes and determine the proportion of direct decays versus those proceeding through an intermediate $^{8}$Be state. The schematics of direct and sequential Hoyle state decay are shown in figure \ref{fig:P1} and this is vital for improving our understanding of nuclear clustering and the processes involved in stellar carbon synthesis.
%\begin{figure}[h]
 %    \centering
   %  \begin{subfigure}[b]{0.45\linewidth}
     %    \centering
     %    \includegraphics[width=\textwidth]{DD_schematic.png}
     %    \caption{}
     %    \label{fig:drift}
    % \end{subfigure}
     % \hspace{30 mm}
   %  \begin{subfigure}[b]{0.45\linewidth}
     %    \centering
     %    \includegraphics[width=\textwidth]{SD_scematic.png}
       %  \caption{}
       %  \label{fig:dff}
    % \end{subfigure}
   %\caption{(a) Schematics of direct decay events, and (b) sequential decay }
     %   \label{fig:P1}
%\end{figure}

%In the AT-TPC, the direct and sequential decay channels can be identified and effectively disentangled by tracking the $\alpha$ particles. It can provide their energy partition following the momentum and energy conservation of the respective decay mechanism. 
In this context, Machine Learning (ML)-based methods, which have shown significant improvements in classification problems \cite{Kriz2012}, may be found useful in designating the relevant tracks and events.
In particle physics experiments with an active target, such as MicroBooNE in the Short Baseline Neutrino program at Fermilab, a liquid argon TPC has incorporated ML methods to classify events and particle tracks \cite{Acc2017}. It has also found applications in nuclear physics experiments in recent times \cite{Kuch2019, Haun2023}. 
%In this work, we have proposed a Neural Network (NN) model for event classification in a prototype AT-TPC, designed at Saha Institute of Nuclear Physics, namlely the SAT-TPC \cite{Das2025}, and identification of the direct and sequential decay channels of the Hoyle state. 
The neural network technique offers a large class of computational models in ML methods, inspired by the way the biological neural network in the human brain processes information. It consists of interconnected layers of nodes, also known as perceptrons. Each of these nodes processes input data, collected from others through non-linear communication, defined by weights. Subsequently, it passes the results, added with approximate biases, to the next layer. The network is trained by adjusting the weights and biases to minimize the error between its output and the recommended result for a given input. Several regularization techniques are used in training the network to prevent overfitting and improve generalization. 
%During training, before showing each batch of examples to the network, a random group of neurons and their connections are temporarily removed. For the next batch, a different set of neurons is removed, and this process continues until the training is finished. By randomly deactivating neurons, the model learns to be more robust and less sensitive to specific features, thus improving its performance on unseen data.
While many different network configurations exist, certain models are found useful in classification problems. Convolutional Neural Network (CNN) is one of the specialised models that is designed to visualize grid-like structured data, such as one-dimensional (1D) time series or 2D grids of pixels of an image \cite{Goo2016}. The network employs a specialized kind of linear operation, called convolution, in place of general matrix
multiplication in at least one of its layers. It is widely used in the field of computer vision for object detection and image recognition and known to involve fewer computational resources. 

We have designed a prototype AT-TPC at Saha Institute of Nuclear Physics, namely the SAT-TPC \cite{Das2025}, to study the Hoyle state decay mechanism and measure the branching ratio of the direct and sequential decay branches. The optimization of design and operational parameters of the SAT-TPC has been carried out with the help of numerical simulation and discussed in \cite{Das2025}. The present work reports the  scheme of data analysis to be adopted for classification of Hoyle state decay events using a CNN model when the SAT-TPC, filled with an active gas mixture Ar + CO$_2$ of volumetric ratio 90:10 at atmospheric pressure, would be subject to 30 MeV $\alpha$-particle beam.     
%In this work, we have implemented a CNN model with Visual Geometry Group (VGG-16) architecture for event classification in a prototype AT-TPC, designed at Saha Institute of Nuclear Physics, namlely the SAT-TPC \cite{Das2025}, which may be useful for identification of Hoyle state decay branches to facilitate the measurement of their individual cross-section. A test case of 30 MeV $\alpha$-particle beam, incident on the active gas mixture Ar + CO$_2$ of volumetric ratio 90:10 at atmospheric pressure, has been subject to numerical simulation. 
The relevant nuclear events, like elastic scattering and the Hoyle state decay following the inelastic scattering of the $\alpha$-projectile from the active target nuclei $^{40}$Ar, $^{12}$C and $^{16}$O, have been generated following numerical simulation of non-relativistic nuclear interaction. The tracks, produced by the scattering and decay products in the horizontal plane containing the beam axis, (referred to as the beam plane in this text), have been reconstructed following the primary ionization of the active gaseous medium and transport of the primary electrons under the action of the electric field in the drift volume of the SAT-TPC. The set of correlated tracks for each event has been considered as a representative 2D image. A large number of such images, representing each class of events, has been produced in order to build the input dataset. A part of the dataset has been used for training and validation of the CNN model while the rest has been utilized for testing the model for its performance in classification of direct and sequential decay branches and their segregation from the elastic scattering events, termed hereafter as background.  An optimization study of the readout granularity of the SAT-TPC, which in principle governs the image resolution, has been performed on the basis of classification efficacy of the CNN model. This approach has the potential to serve as an automated analysis framework for tagging and separating actual experimental data.

%For the present work, the CNN model has been opted for classification of the Hoyle state decay channels in presence of other reaction channels, considered here as the backgound. For this purpose, the tracks of the events produced via $\alpha$ with energy 30 MeV bombarding on active gas mixture $Ar + CO_2$ of volumetric ratio of 90:10 at atmospheric pressure have been simulated and analysed. 

The article has been organized in the following manner. The workflow of the present numerical study starting from the generation of the scattering and decay events to the training of the CNN model has been described in section \ref{NumStud}. All the stages of the numerical simulation have been briefly discussed in several sub-sections of the same. The results of the implementation of the CNN model in classification of nuclear events have been presented in section \ref{PerAnal}. The article concludes with section \ref{Con}. 

\section{Numerical Study}
\label{NumStud}
\begin{figure}[h!]
 \centering
     \includegraphics[scale=0.5]{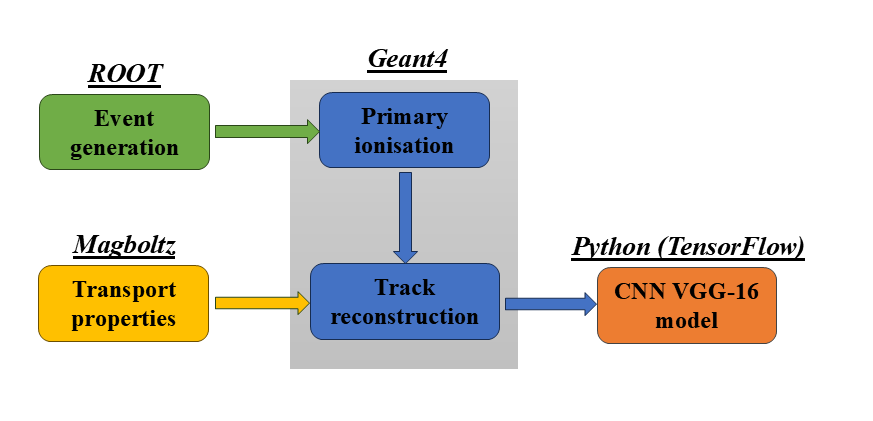}
 \caption{Flowchart of the numerical model}
     \label{flowchart}
   \end{figure}

The flowchart of the numerical model, followed in this work to address the classification problem of the Hoyle state decay branches in the case of a 30 MeV $\alpha$-projectile interacting with an Ar + CO$_2$ (90:10) gas mixture, is illustrated in figure \ref{flowchart}. The relevant software toolkits associated with each stage to perform the numerical work are mentioned as well. A brief discussion on individual stages has been provided in the following sub-sections.

\subsection{Event Generation}
\label{EveGen}
In the interaction of 30 MeV $\alpha$-projectile with the active gas mixture Ar + CO$_2$, five cases of nuclear events have been considered in this work.  These are the elastic scattering of the $\alpha$-projectile with the active targets $^{40}$Ar, $^{12}$C, $^{16}$O nuclei and the inelastic scattering of the $\alpha$-particle with $^{12}$C only, followed by direct and sequential decay of the Hoyle state, each producing three $\alpha$-particles. 

\begin{figure}[h]
    \centering
    \begin{subfigure}[b]{\textwidth}
        \centering
        \includegraphics[width=0.45\textwidth]{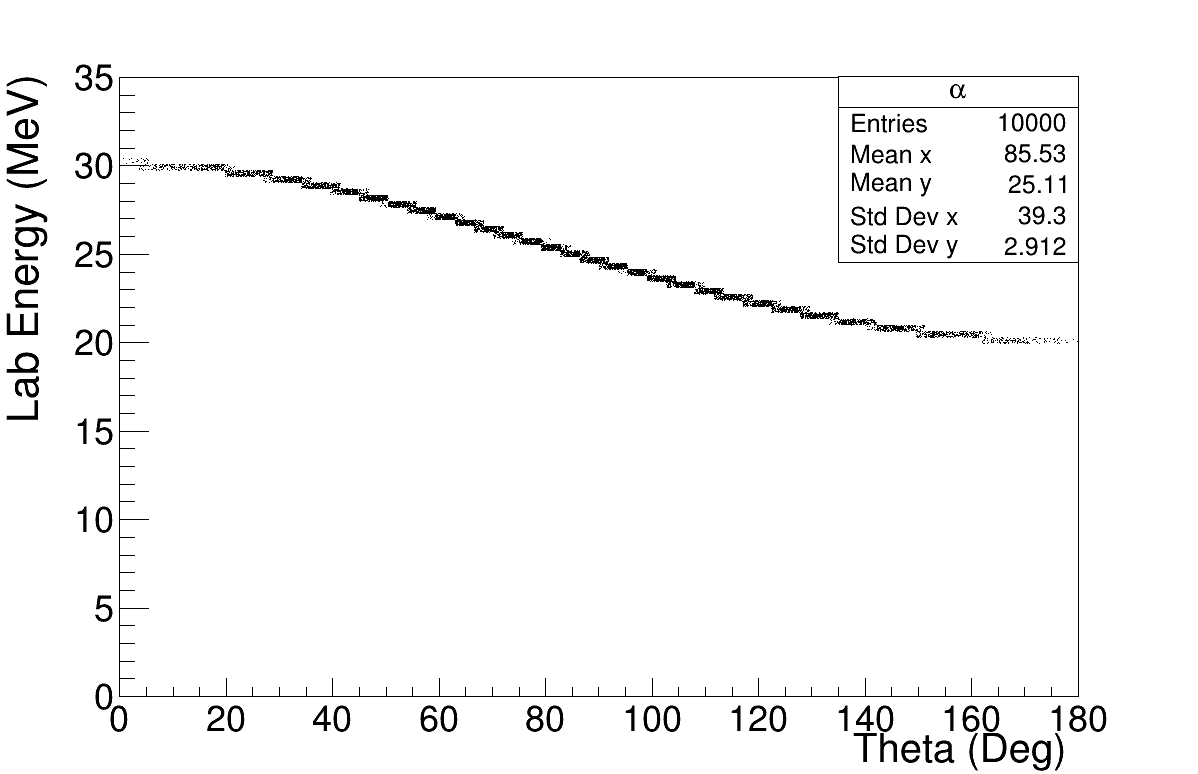}
        \includegraphics[width=0.45\textwidth]{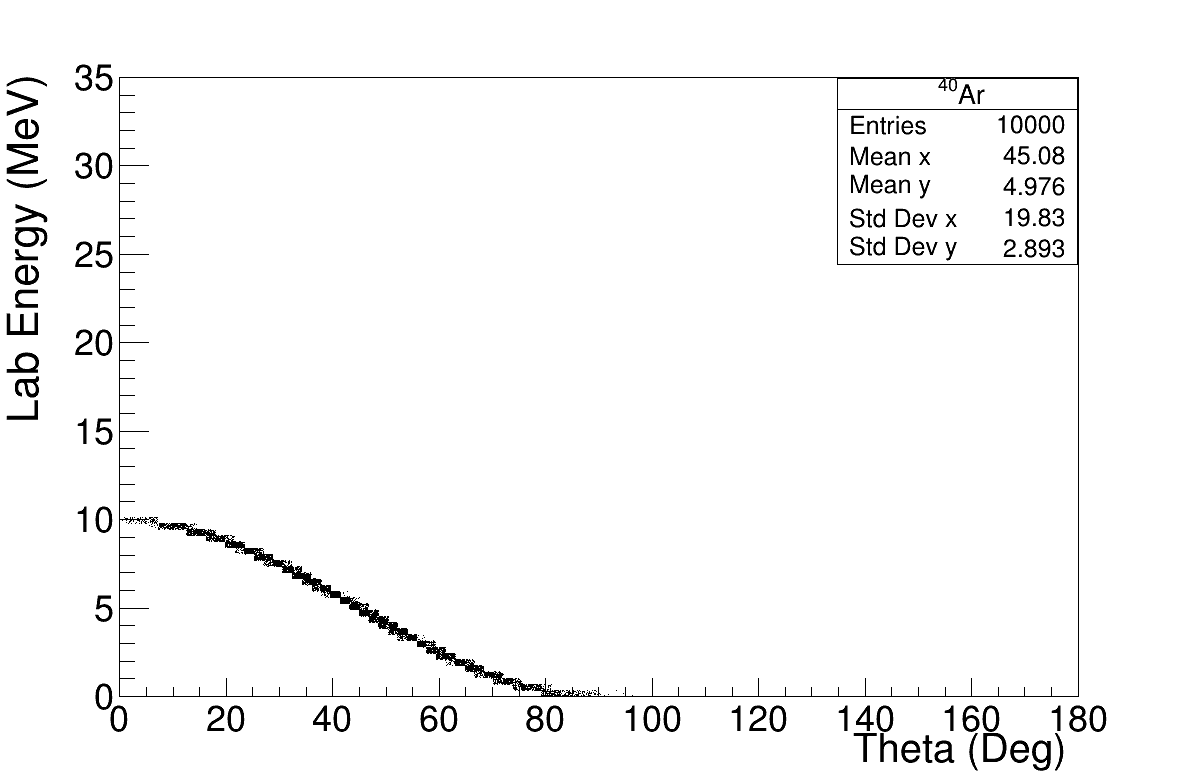}
        \caption{$\alpha$ + $^{40}$Ar}
     \end{subfigure}
     \begin{subfigure}[b]{\textwidth}
       \centering
        \includegraphics[width=0.45\textwidth]{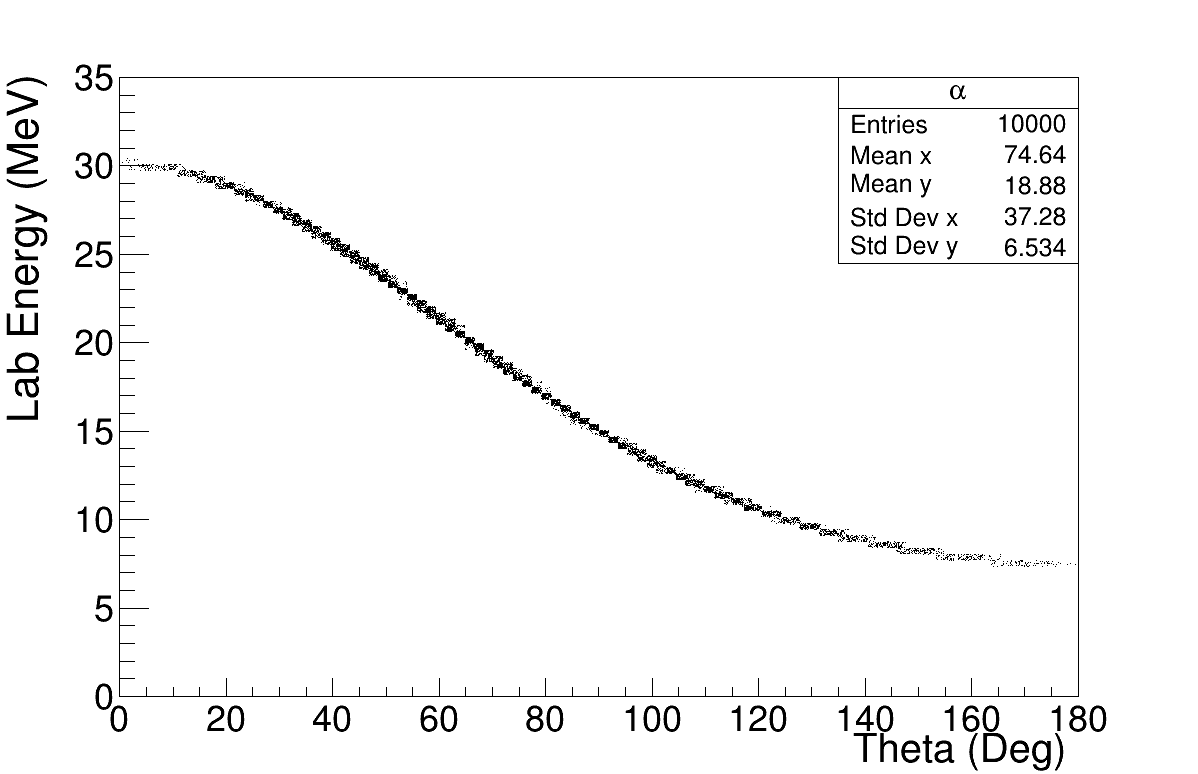}
        \includegraphics[width=0.45\textwidth]{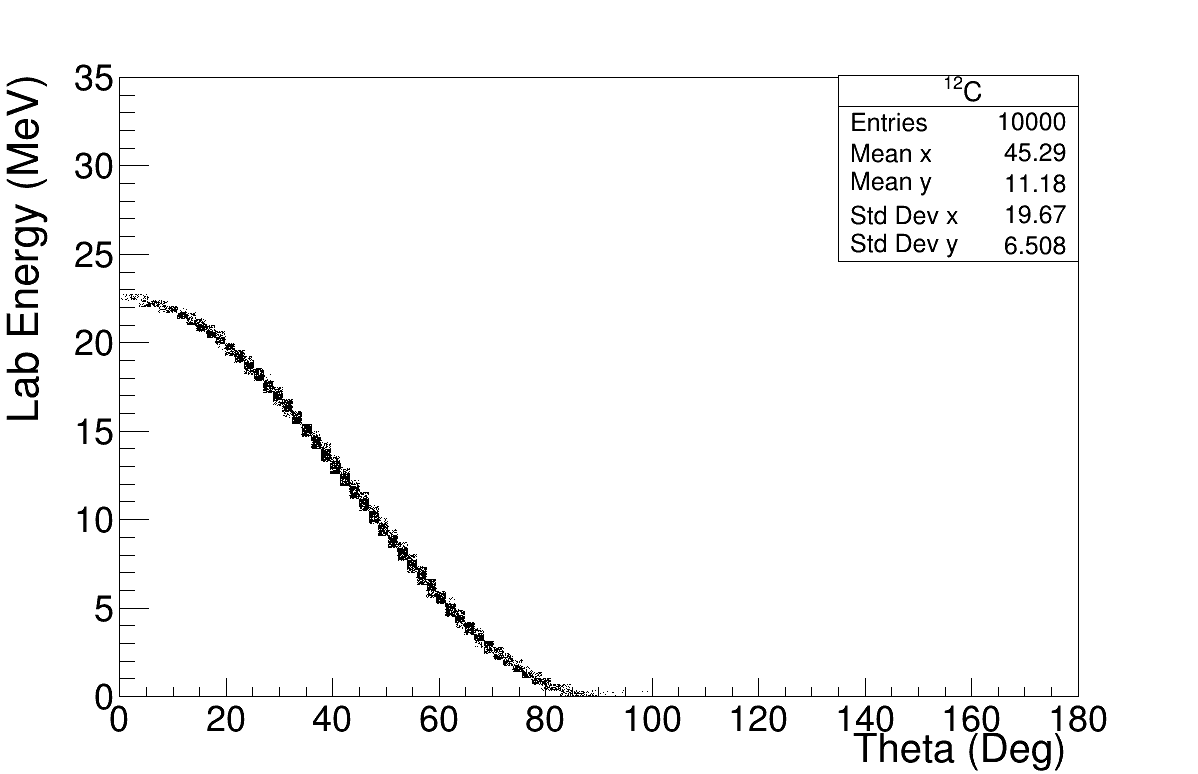}
        \caption{$\alpha$ + $^{12}$C}
     \end{subfigure}
     \begin{subfigure}[b]{\textwidth}
       \centering
        \includegraphics[width=0.45\textwidth]{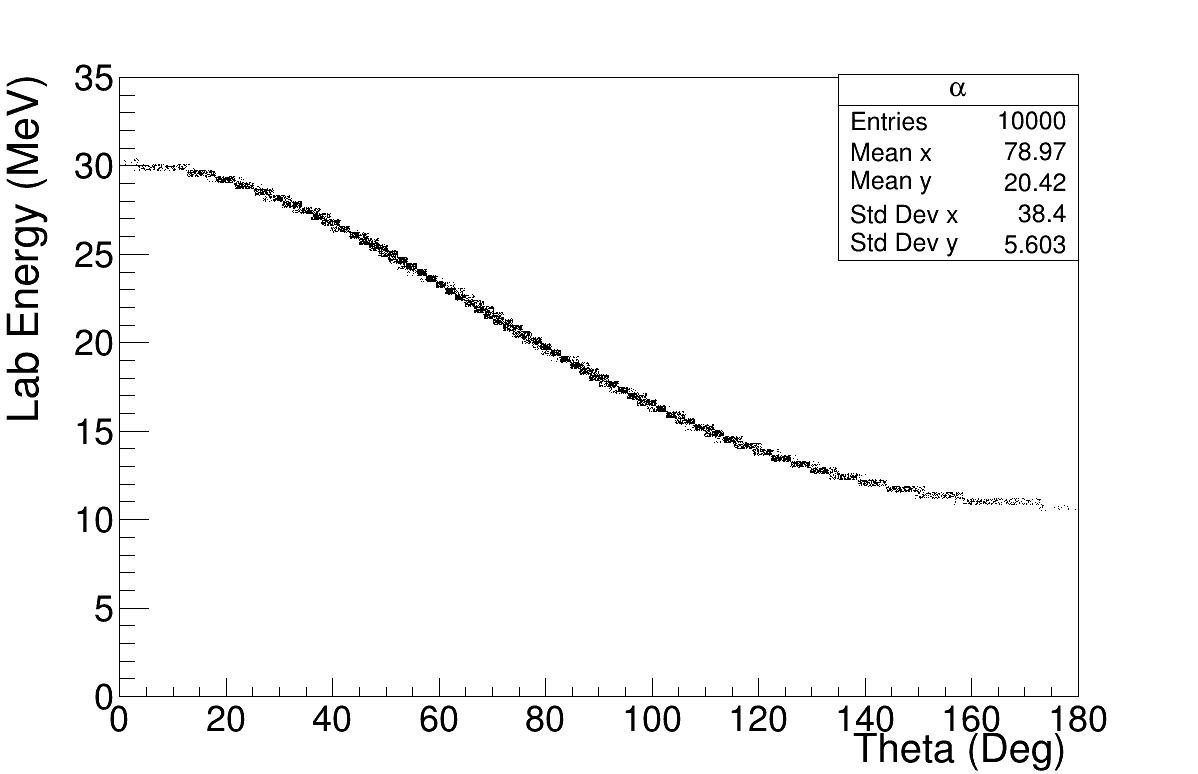}
        \includegraphics[width=0.45\textwidth]{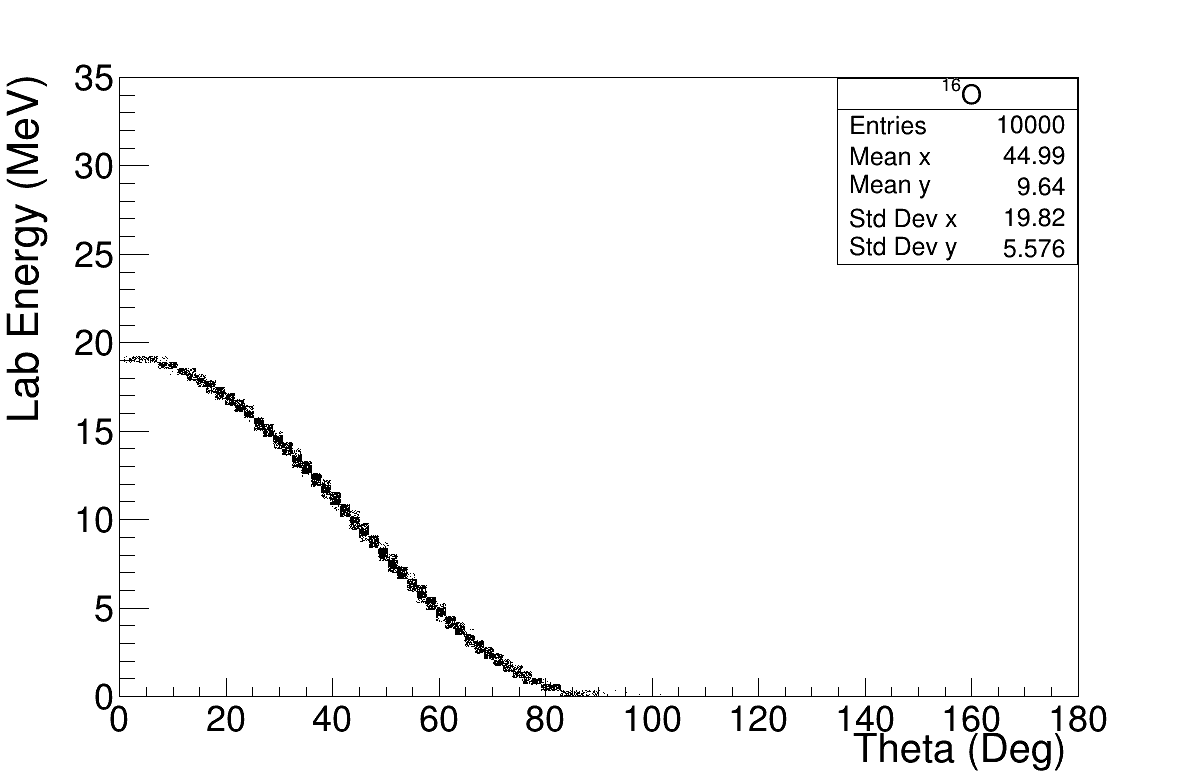}
        \caption{$\alpha$ + $^{16}$O}
     \end{subfigure}
\caption{Correlated angle and energy of scattering products in lab frame for elastic scattering cases: (a) $\alpha$ + $^{40}$Ar, (b) $\alpha$ + $^{12}$C, and (c) $\alpha$ + $^{16}$O}
 \label{EG_scattered}
\end{figure}

An event-by-event Monte Carlo simulation of the elastic scattering has been developed in ROOT framework \cite{Bru1997} using the TGenPhaseSpace class following non-relativistic kinematics for all possible angles in the center of mass (c.m.) frame.
The energy and corresponding angle of the scattering products have been obtained by boost in the laboratory (lab) frame. The correlation of the angle, $\theta$, and the energy in the lab frame, $E_{lab}$ of the products of the three cases of elastic scattering $\alpha$ + $^{40}$Ar, $\alpha$ + $^{12}$C, and $\alpha$ + $^{16}$O, are plotted in the figure \ref{EG_scattered}, respectively. These data have been used to produce the primary ionization tracks of the corresponding scattering products, as discussed in sub-section \ref{TrkGen}.

%The tracks created by these products through primary ionization of the active target gaseous molecules have been simulated using Geant4 \cite{Ago2003}. From the transport properties of the electrons in the primary tracks, as determined by Magboltz \cite{Bia1999}, the diffusion parameters have been convoluted to reconstruct the 2D tracks projected on the SAT-TPC readout plane. The event kinematics as represented by the correlated tracks has been recorded as image for training and validation of the CNN model. The CNN model with the VGG-16 architecture has been trained and validated with a part of the image dataset produced for the events. The rest part has been used for studying the performance of the model in distinguishing the Hoyle state decay from the elastic scattering events and further classification of direct and sequential decay branches. 
%The corresponding 3D tracks produced by the scattering and decay products of each event have been simulated considering the primary ionization of the gaseous molecules. The tracks associated with each event have been convoluted by electron diffusion parameters, have been used for training and validation of the proposed CNN model. 

\begin{figure}[h]
    \centering
    \begin{subfigure}[b]{\textwidth}
         \centering
         {\includegraphics[width=0.3\textwidth]{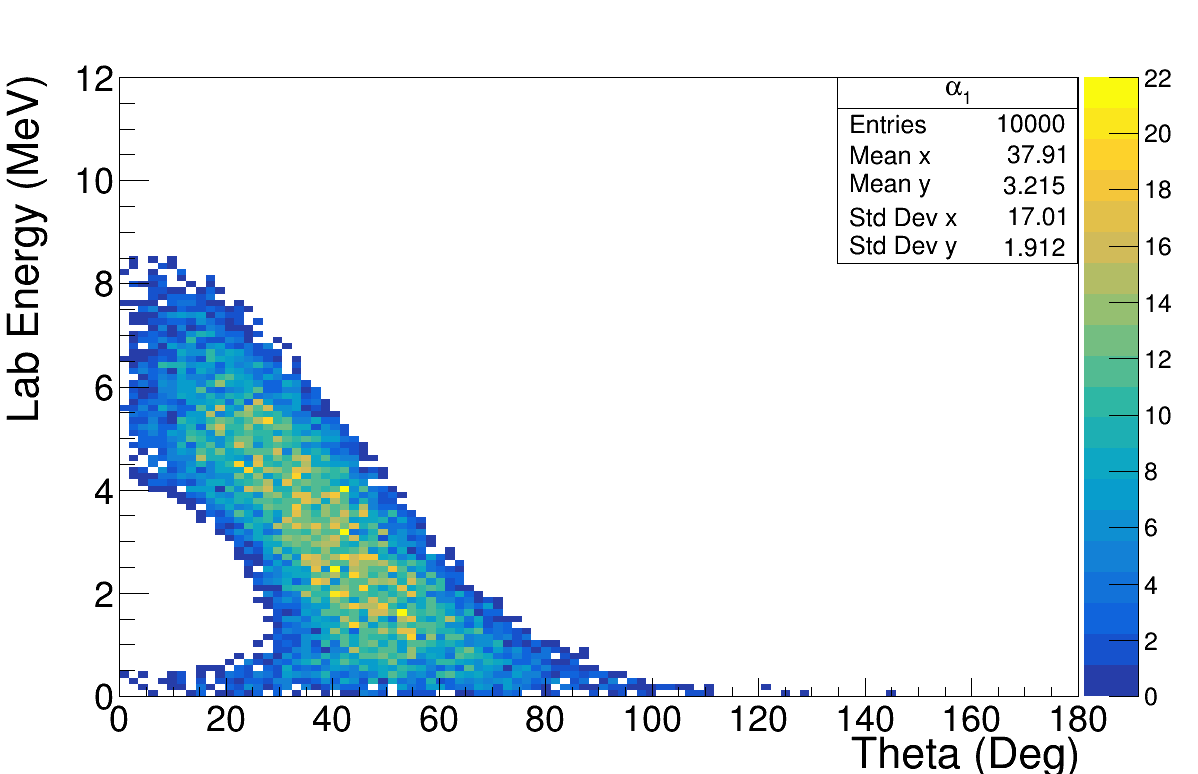}}
         {\includegraphics[width=0.3\textwidth]{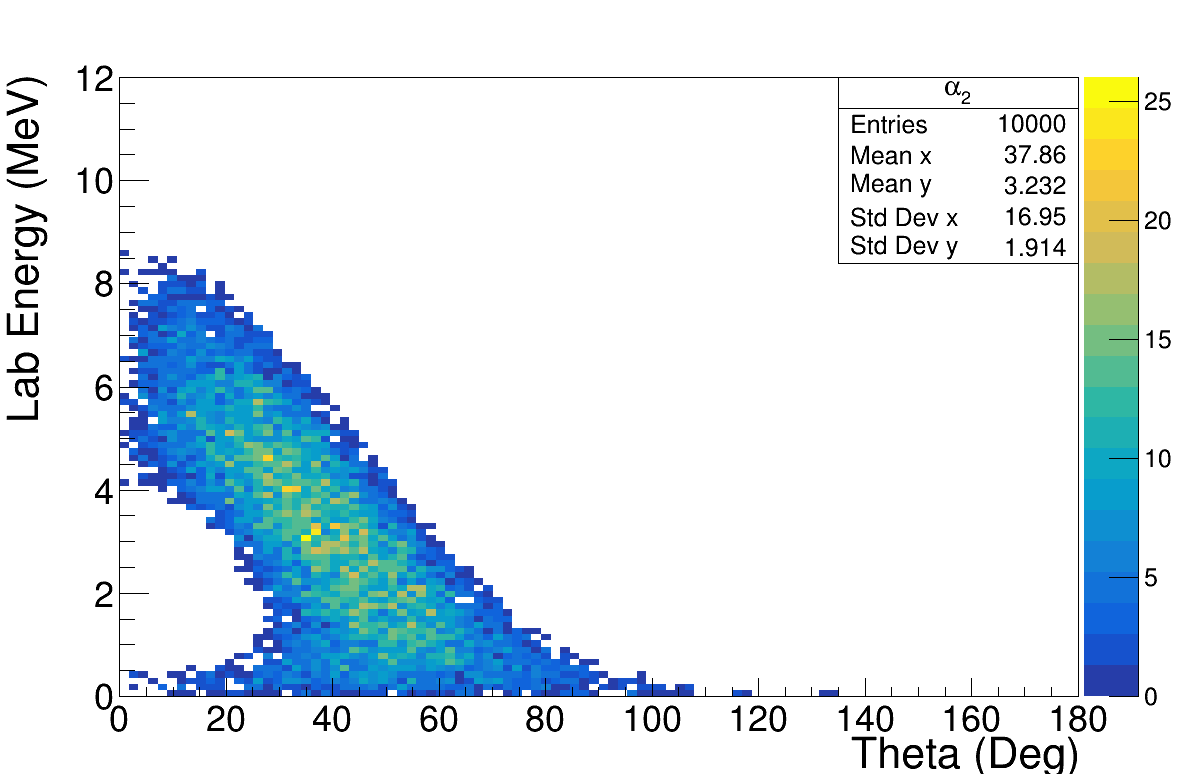}}
        {\includegraphics[width=0.3\textwidth]{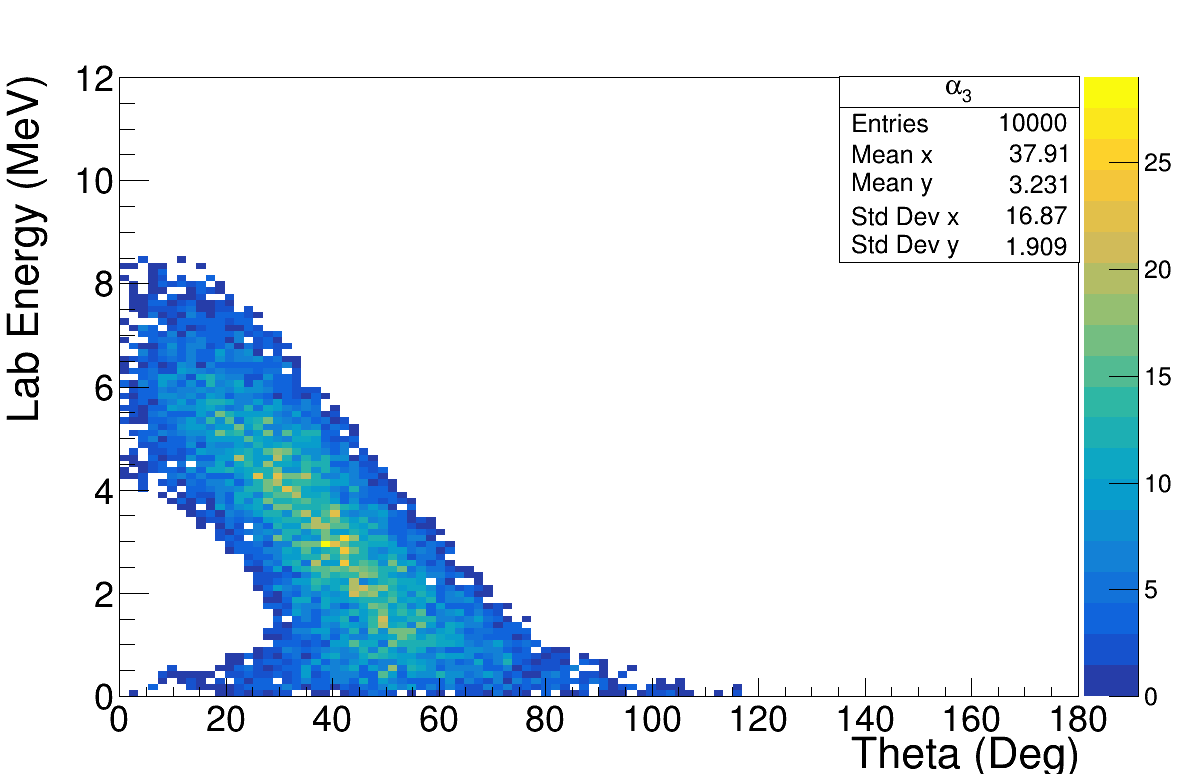}}
        \caption{Direct decay}
   \end{subfigure}
   \begin{subfigure}[b]{\textwidth}
       \centering
       {\includegraphics[width=0.3\textwidth]{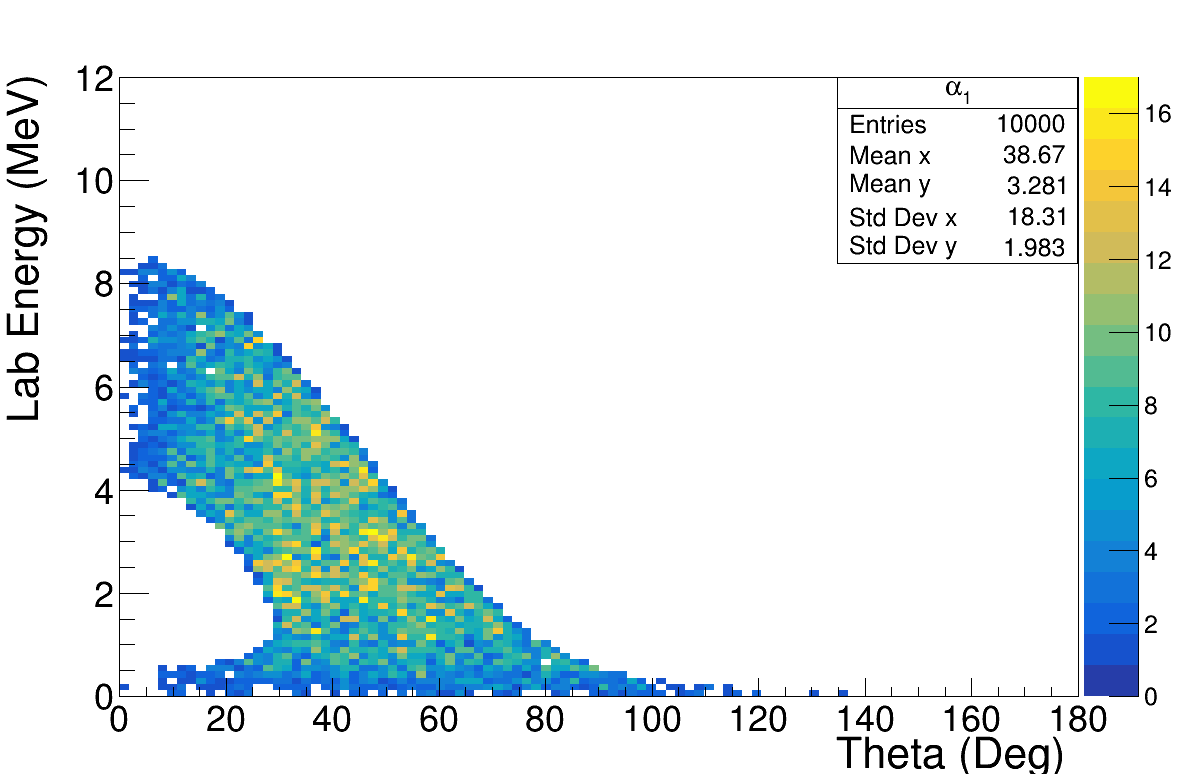}}
       {\includegraphics[width=0.3\textwidth]{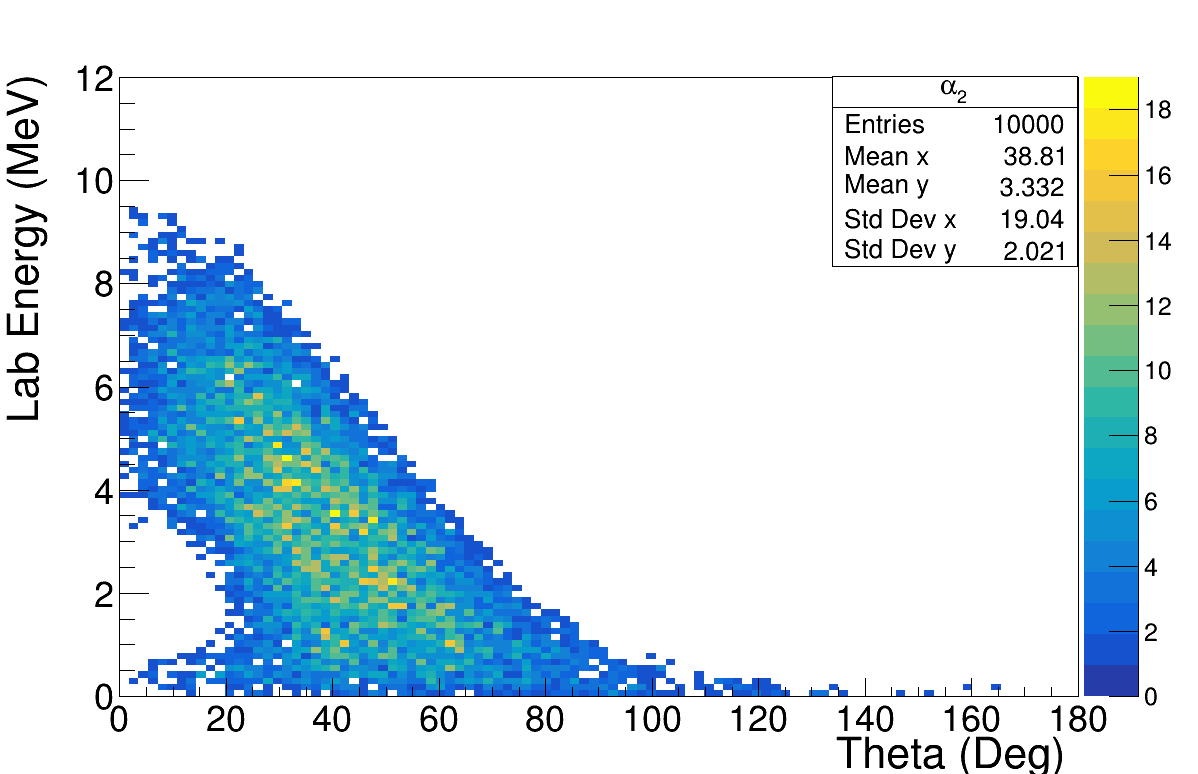}}
       {\includegraphics[width=0.3\textwidth]{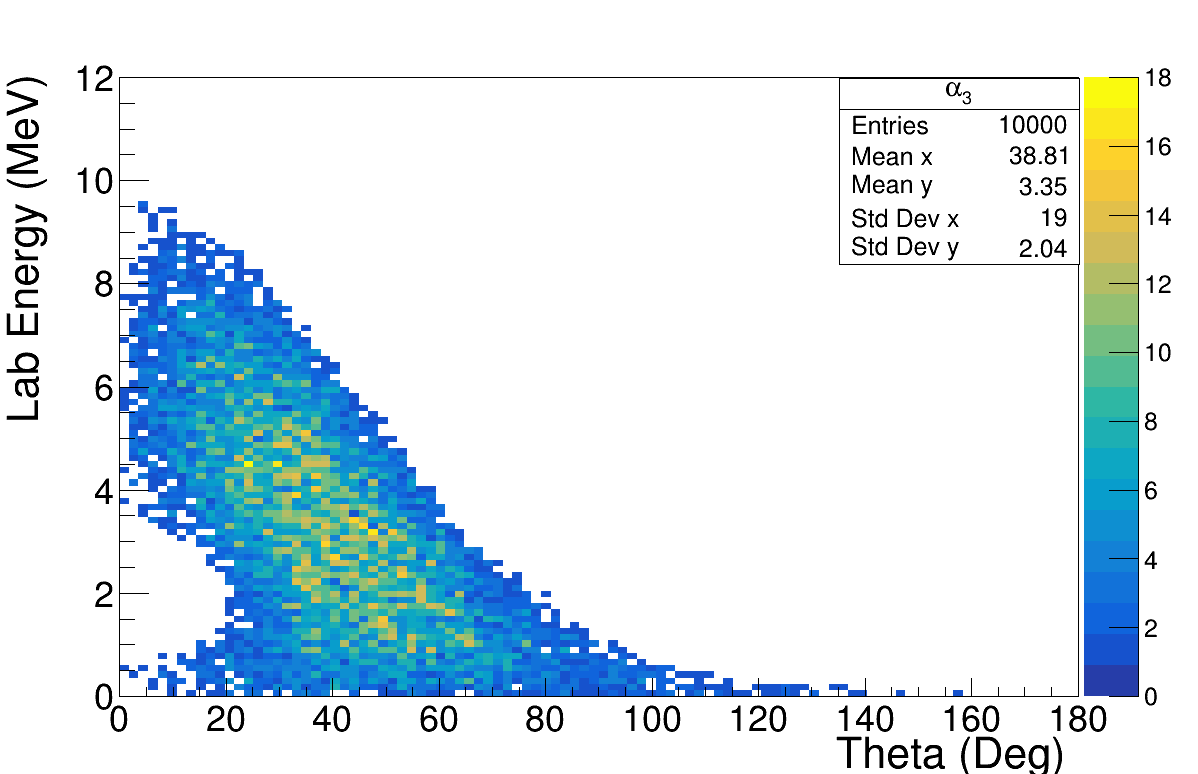}} 
       \caption{Sequential decay}
       \end{subfigure}
 \caption{Correlated angle $\theta$ and energy $E_{lab}$ in the lab frame of three $\alpha$-particles in Hoyle state decay branches: (a) Direct decay, and (b) Sequential decay}
 \label{EG_decay}
\end{figure}

We have investigated the dynamics of the Hoyle state focusing on the excitation of the $^{12}$C nucleus through inelastic scattering of an $\alpha$-particle followed by the decay mechanisms. The direct decay of the $^{12}$C$^{\ast}$ as well as its sequential decay via $^{8}$Be + $\alpha$, all leading to triple-$\alpha$ products, have been modeled using a phase-space generator. The resulting energy, momentum and angular distribution of the $\alpha$-particles have been obtained after applying a relativistic boost to the decay products of the Hoyle state, generated in the c.m. frame. For the direct decay of the Hoyle state, the triple-$\alpha$ decay has been boosted with the four-momentum vector of $^{12}$C.  
In case of sequential decay, it has been done in two steps. In the first step, $\alpha$ and $^{8}$Be have been generated and boosted in the direction of $^{12}$C. In the next step, the two $\alpha$-particles, generated from $^{8}$Be, have been boosted in the direction of $^{8}$Be. The four-momentum, obtained for all the nuclei in each event, has been used to produce the corresponding tracks, as detailed in the next sub-section \ref{TrkGen}. These parameters have been recorded in histograms and the 2D correlation plots of the angle, $\theta$, and the energy $E_{lab}$ in the lab frame, are shown in figure \ref{EG_decay}.
 
 \subsection{Track Reconstruction}
\label{TrkGen}
\begin{figure}[h]
     \centering
     \begin{subfigure}[b]{0.45\linewidth}
         \centering
         \includegraphics[width=\textwidth]{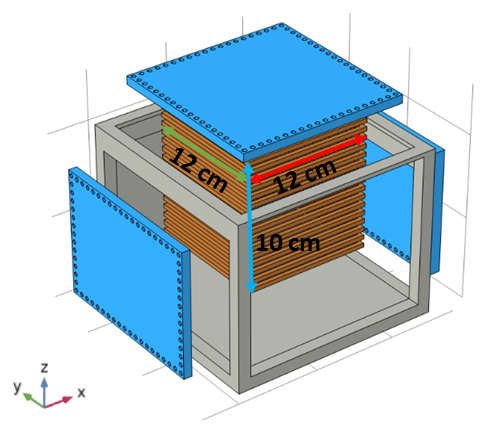}
         \caption{}
         \label{ED}
     \end{subfigure}
     % \hspace{30 mm}
     \begin{subfigure}[b]{0.5\linewidth}
         \centering
         \includegraphics[width=\textwidth]{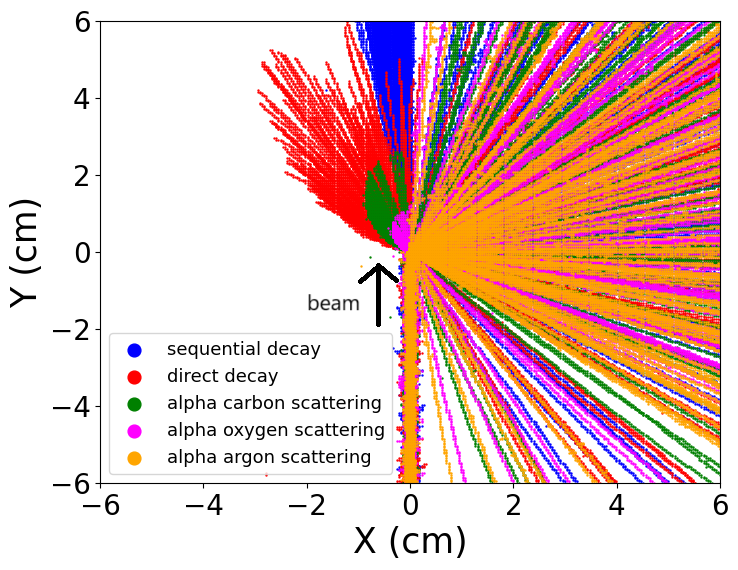}
         \caption{}
         \label{events}
     \end{subfigure}
   \caption{(a) Schematic design of the SAT-TPC, and (b) correlated primary tracks of 100 events of each class, as simulated in Geant4} 

        \label{fig}
\end{figure}

The geometry of the SAT-TPC, depicted in figure \ref{ED}, has been modeled in Geant4 \cite{Ago2003}. The uniform drift field of 500 V/cm in the active region of the TPC along the Z-axis has been considered for this work.

The primary ionization in the active gas volume, caused by the scattering and decay products which are ejected with their respective energies along their trajectories within the active gaseous volume of the SAT-TPC, has been simulated with Geant4. To incorporate the low-energy electromagnetic processes, the physics lists Penelope, Livermore, and Photo Absorption and Ionization (PAI) in Geant4 have been taken into account. Only those events with their product tracks, lying on the beam plane (XY) in the active volume of the SAT-TPC, have been taken into account to simplify the calculation. For each case of elastic scattering and Hoyle state decay branches,1000 such events have been considered (i.e. 5000 in total). The correlated primary tracks of all the cases of the nuclear events, as obtained from Geant4, have been collectively illustrated in figure \ref{events} for 100 events in total with 20 for each case. 

\begin{figure}[h]
    \centering
    \begin{subfigure}[b]{\textwidth}
       \centering
        {\includegraphics[width=0.3\textwidth]{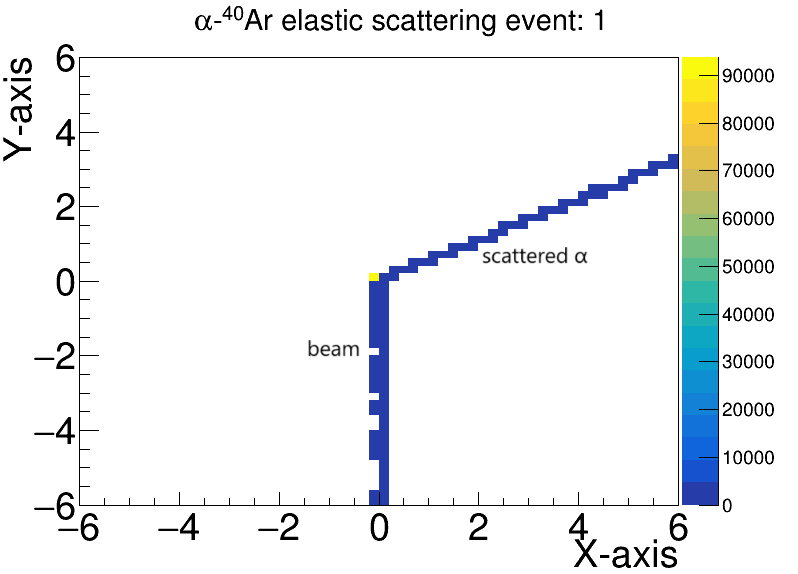}}
        {\includegraphics[width=0.3\textwidth]{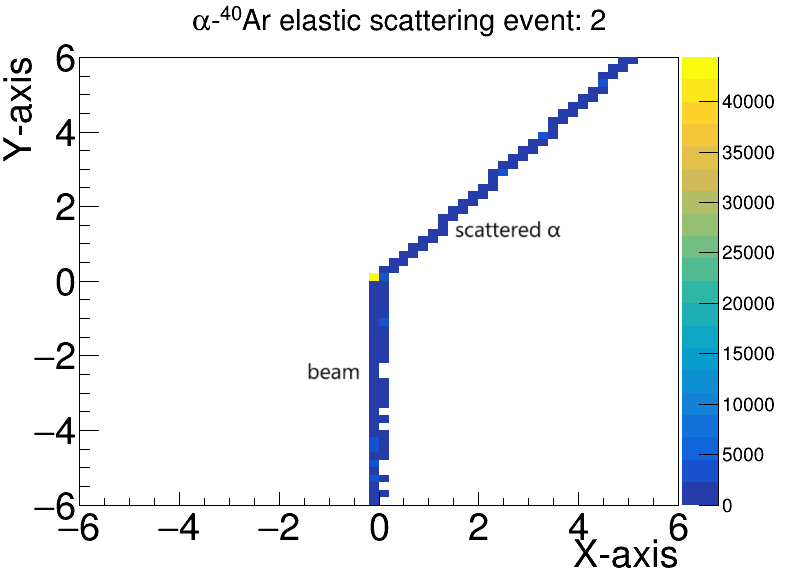}}
        {\includegraphics[width=0.3\textwidth]{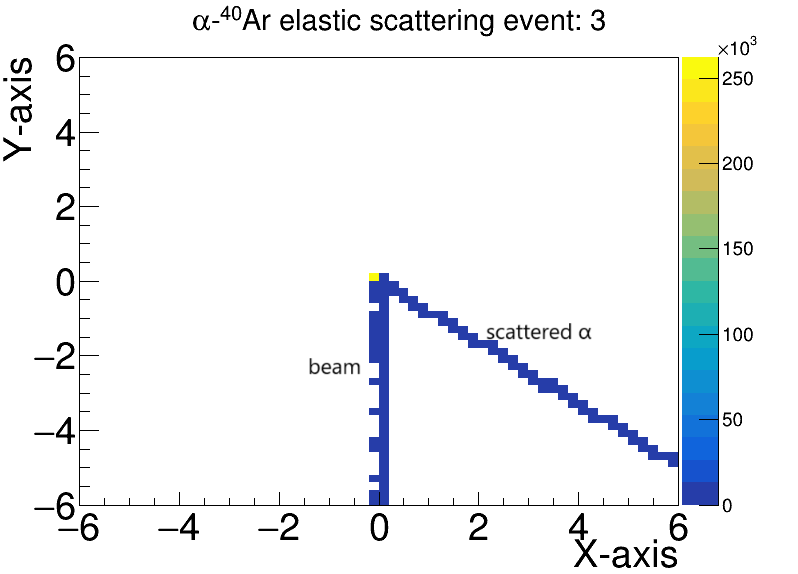}}
       \caption{$\alpha$ + $^{40}$Ar}
     \end{subfigure}
     \begin{subfigure}[b]{\textwidth}
       \centering
       {\includegraphics[width=0.3\textwidth]{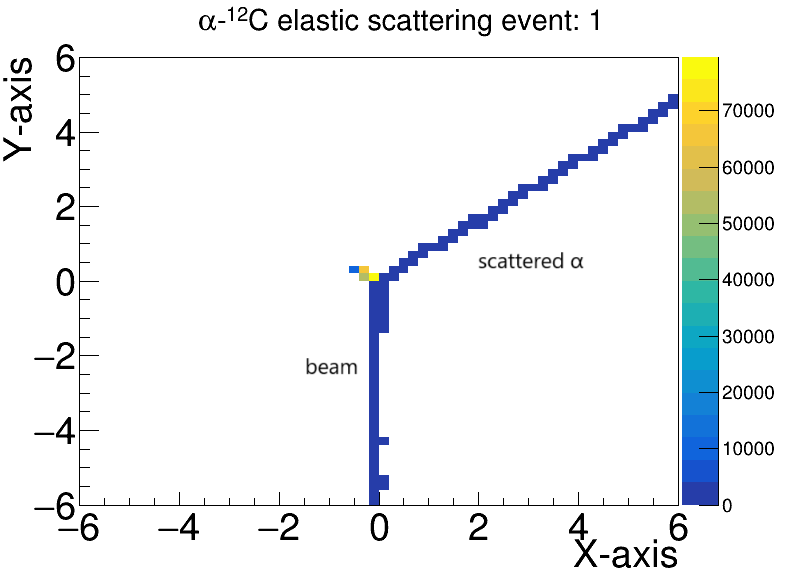}}
       {\includegraphics[width=0.3\textwidth]{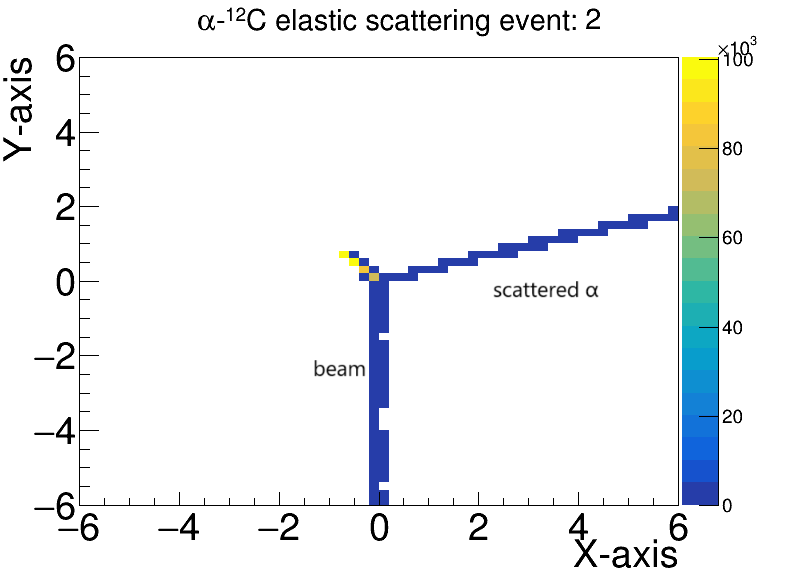}}
       {\includegraphics[width=0.3\textwidth]{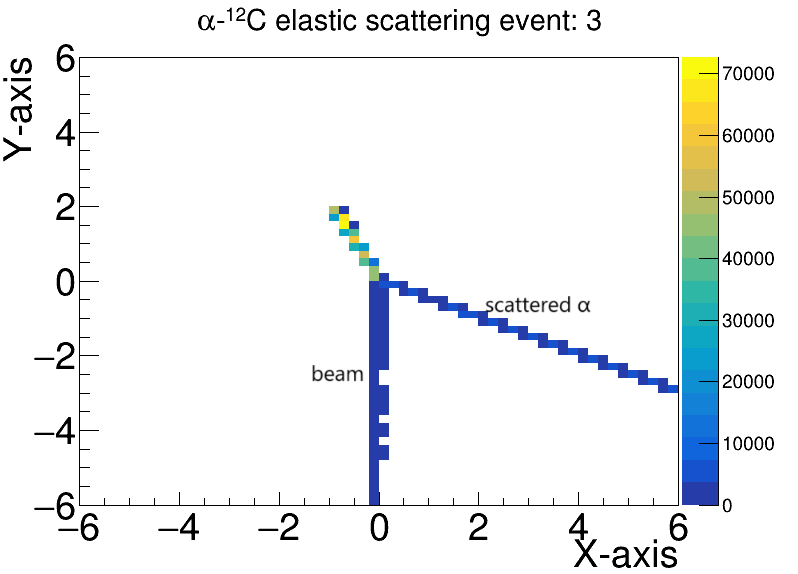}}
       \caption{$\alpha$ + $^{12}$C}
     \end{subfigure}
     \begin{subfigure}[b]{\textwidth}
     \centering
     {\includegraphics[width=0.3\textwidth]{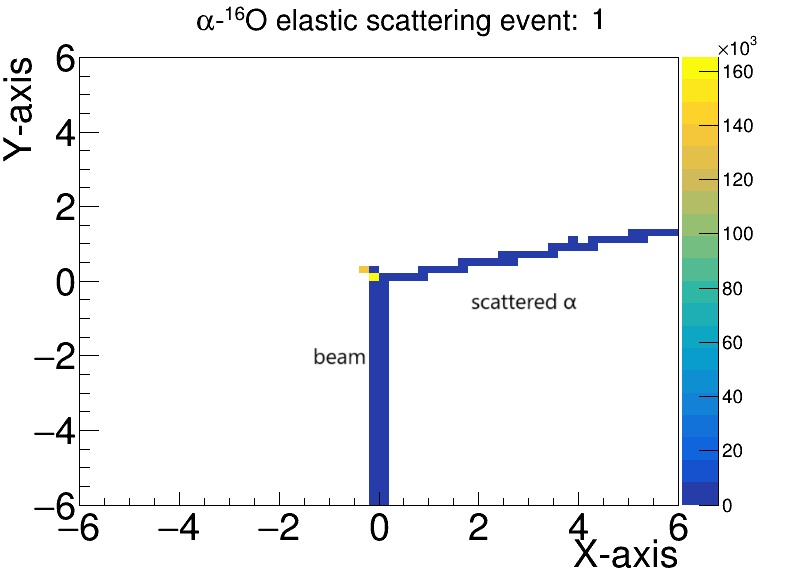}}
     {\includegraphics[width=0.3\textwidth]{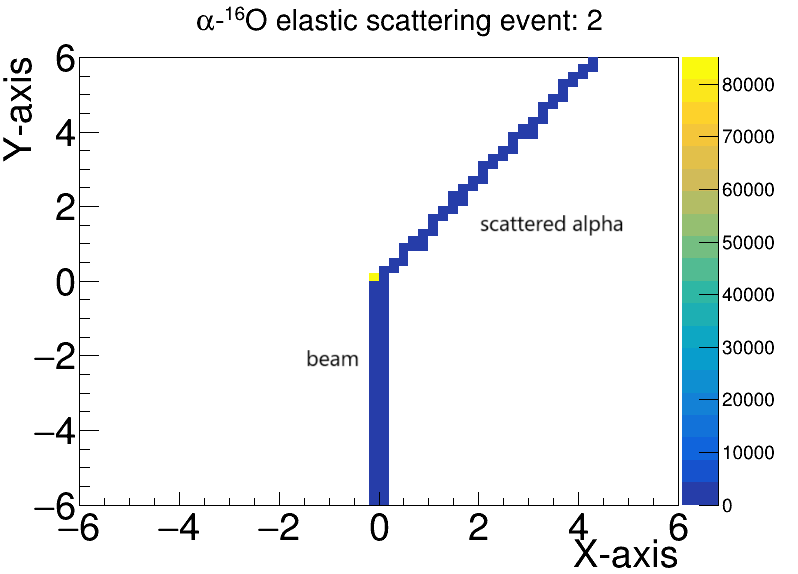}}
     {\includegraphics[width=0.3\textwidth]{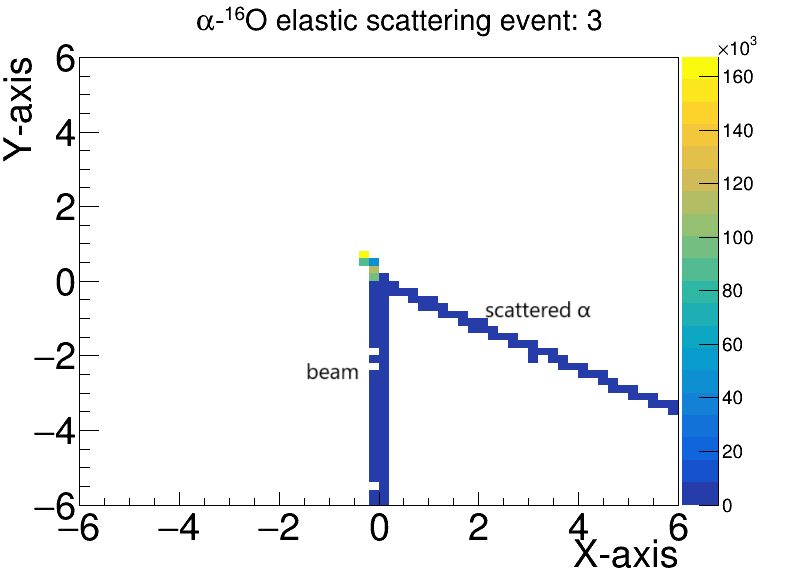}} 
    \caption{$\alpha$ + $^{16}$O}
     \end{subfigure}
    \caption{Reconstructed tracks on the SAT-TPC readout plane of the elastic scattering cases: (a) $\alpha$ + $^{40}$Ar, (b) $\alpha$ + $^{12}$C, and (c) $\alpha$ + $^{16}$O}
    \label{scat}
\end{figure}

\begin{figure}[h]
    \centering
    \begin{subfigure}[b]{\textwidth}
       \centering
       {\includegraphics[width=0.3\textwidth]{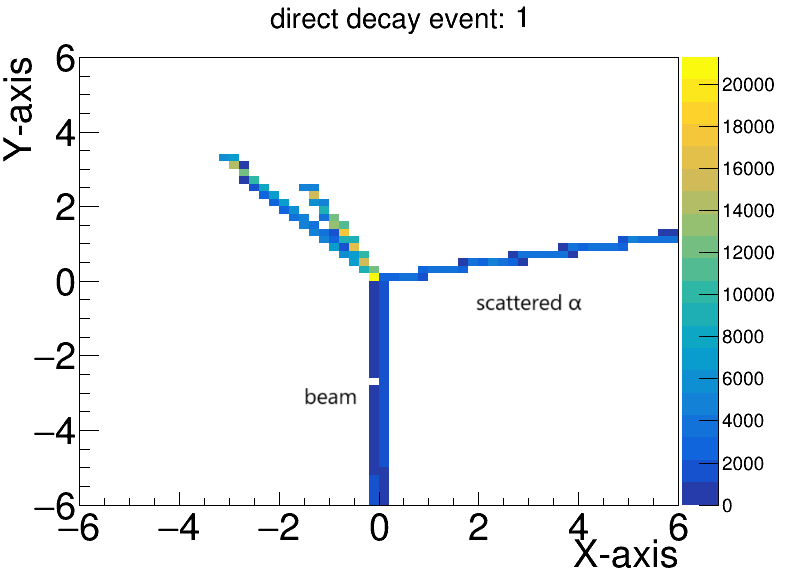}}
       {\includegraphics[width=0.3\textwidth]{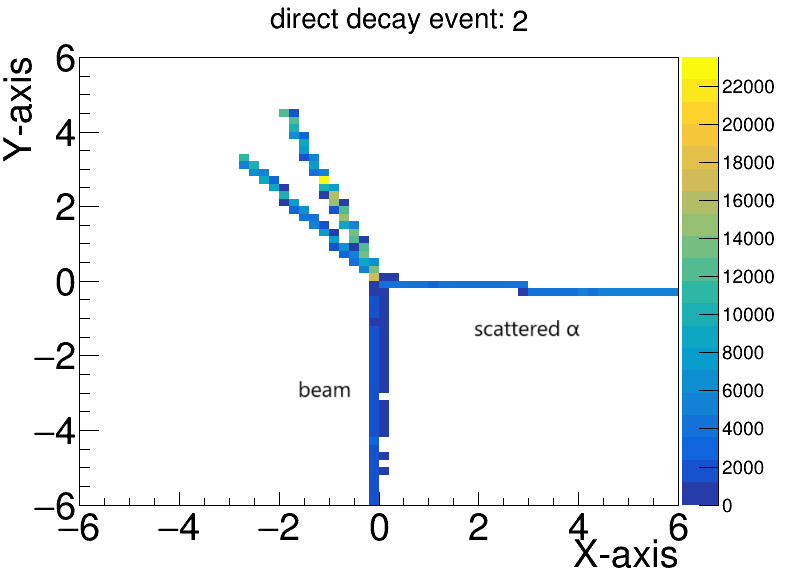}}
       {\includegraphics[width=0.3\textwidth]{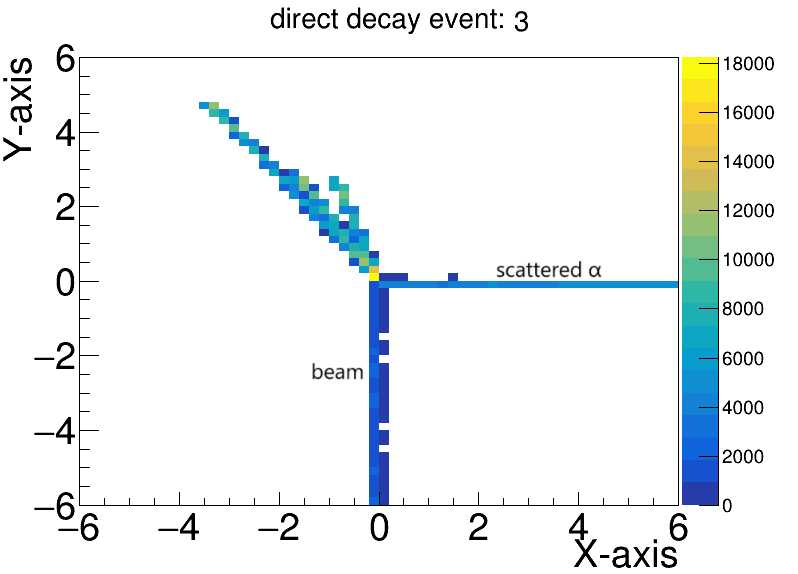}} 
      \caption{Direct decay}
    \end{subfigure}
    \begin{subfigure}[b]{\textwidth}
    \centering
    {\includegraphics[width=0.3\textwidth]{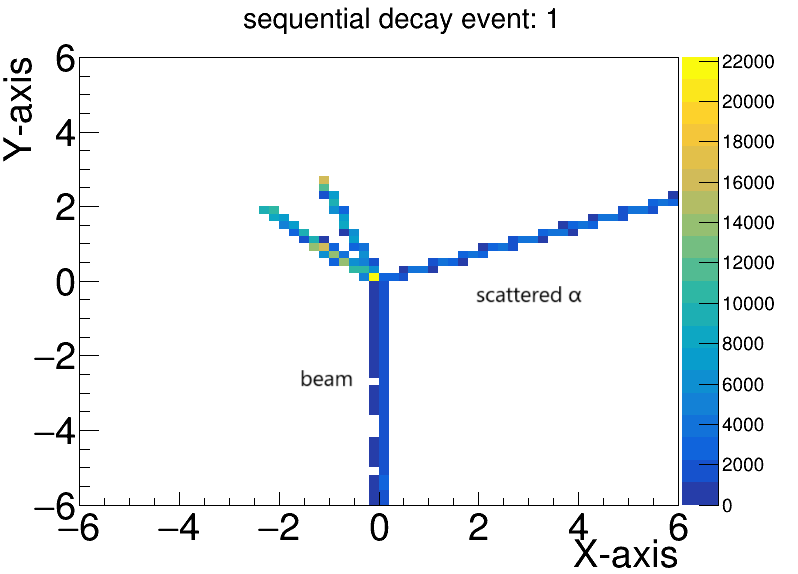}}
    {\includegraphics[width=0.3\textwidth]{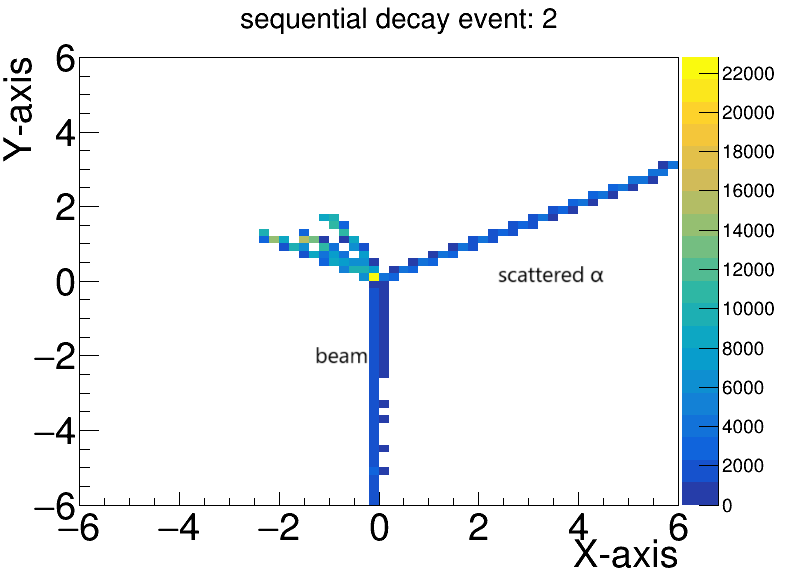}}
    {\includegraphics[width=0.3\textwidth]{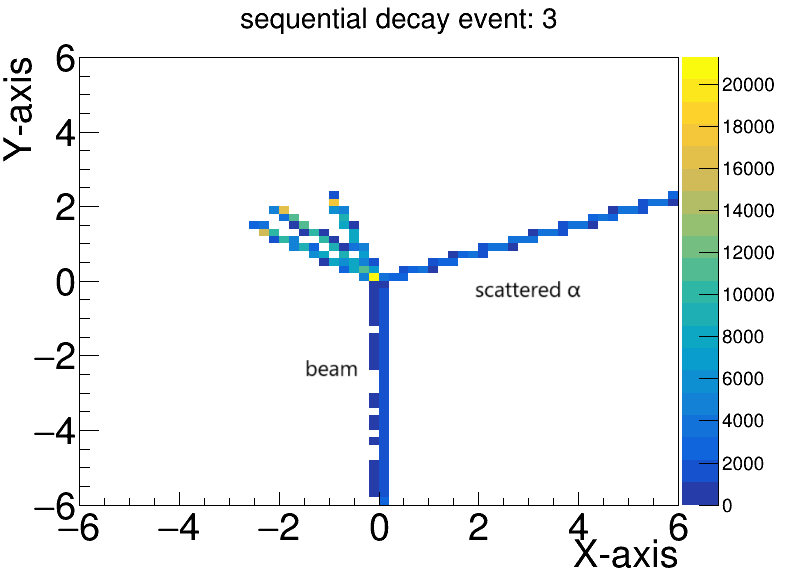}} 
    \caption{Sequential decay}
    \end{subfigure}
\caption{Reconstructed tracks on the SAT-TPC readout plane of the Hoyle state decay branches: (a) direct decay, and (b) sequential decay}
 \label{decay}
\end{figure}

In the next step, the primary tracks for each event have been projected on the readout plane of the SAT-TPC with sensitive area 12 cm $\times$ 12 cm, consisting of 60 $\times$ 60 square pixels.  No temporal evolution of the primary tracks through the drift volume of the SAT-TPC has been considered as the simulation of the same for 5000 events would be computationally expensive. However, the effect of diffusion has been convoluted in the primary tracks using a Gaussian profile with a variance, equivalent to the transverse diffusion of the electrons in the given electric field of the SAT-TPC, as produced by the Magboltz toolkit \cite{Bia1999} in the Garfield++ framework \cite{Vee1998}. A detailed discussion on the temporal evolution of the primary tracks may be found in our previous work \cite{Das2025}. 
Three typical examples of reconstructed tracks, projected on the readout plane with 60 $\times$ 60 pixels, each for three cases of elastic scattering, are illustrated in figure \ref{scat}. It can be observed that the reconstructed track of the scattered $^{40}$Ar is too short in comparison to that of the scattered $^{12}$C and $^{16}$O nuclei, being the heaviest among the three products. Similarly, three examples of the reconstructed tracks for each of the direct and sequential decay events, following convolution and projection of the primary tracks on the readout plane, are shown in figure \ref{decay}. These reconstructed tracks corresponding to each class of nuclear events have been recorded as input image data for training and validation, followed by testing of the performance of the CNN model in classification of Hoyle state decay branches.

\subsection{CNN Model}
% A schematic diagram of different stages of a CNN model is depicted in figure \ref{fig:CNN}.

In the first step, a convolution filter having a smaller width and height and the same depth as the input volume, is used to extract relevant features from the input data, such as edges, corners, shapes, etc. The filter slides over the input grid performing element-wise multiplication and summing to produce 2D results without changing the dimension. The results, produced by each filter, are stacked together, and as a result, we get an output volume, having a depth equal to the number of filters. In the next step, an activation layer adds nonlinearity to the network by applying an element-wise activation function to the output of the convolution layer, keeping the data volume unchanged. It allows the network to learn more complex patterns and a few examples of the same are ReLU (Rectified Linear Unit), Tanh, etc. After this layer, a pooling function is used to down-sample the input data of the previous layer, which in turn helps in reducing computational load, and thus improving the efficiency of the network. It also helps to minimize overfitting and reduce dimensionality. Additionally, dropout is used to reduce the over-reliance on specific neurons. Common types of pooling functions include max pooling and average pooling. It is followed by the flattening of the resulting feature map into a 1D vector.  In the fully connected layer, the final classification or regression task is performed on the basis of the learned features. At the end stage, the output from the fully connected layer is fed into a logistic function to convert the output into the probability score of each class, (the elastic scattering and the Hoyle state decay events).
 % \begin{figure}[h!]
 % \centering
 %     \includegraphics[scale=0.45]{CNN.png}
 % \caption{Neural network}
 %     \label{fig:CNN}
 %   \end{figure} 

\begin{figure}[h!]
 \centering
     \includegraphics[scale=0.5]{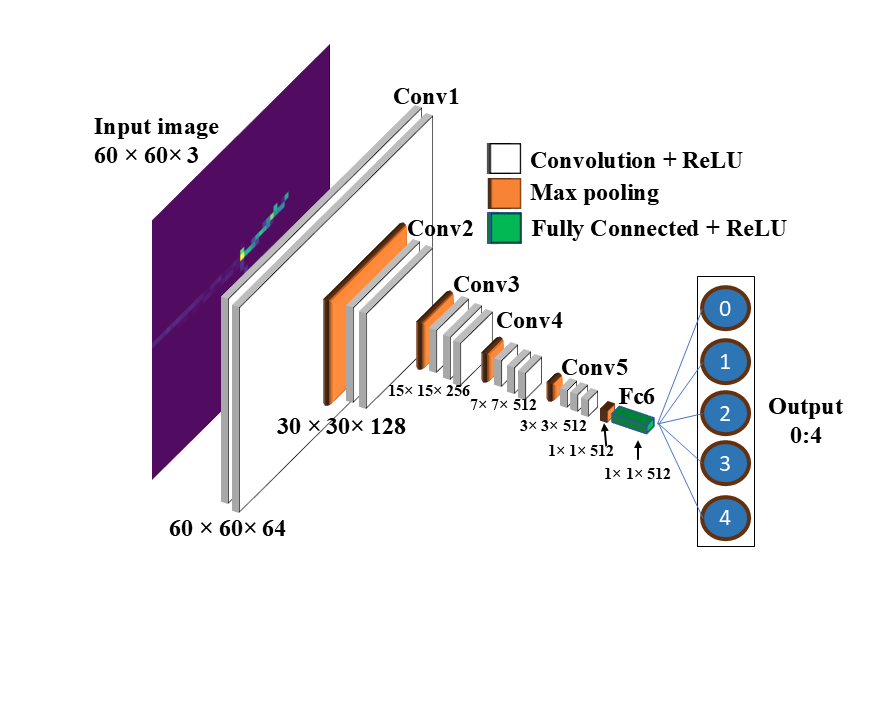}
 \caption{Scheme of VGG-16 architecture in the present CNN model}
     \label{fig:m_vgg}
   \end{figure}

In the present work, we have used the VGG-16 (Visual Geometry Group) architecture \cite{Sim2014} to design the CNN model which is trained on the ImageNet repository \cite{Den2009}. It is a widely popular and highly efficient CNN architecture for image recognition and very competitive even today, compared to other architectures, such as ResNet or DenseNet. The VGG-16 is well known for its simple structure and distinguished by the application of a small convolution filter of size 3 $\times$ 3 pixels. The scheme of the said architecture, as has been used in the current work for classification of the nuclear events, is illustrated in figure \ref{fig:m_vgg}. The input shape for the CNN model is [channels, height, width] which is equivalent to the 2D color images with red, green and blue values in the third dimension of the image pixels. For our study, we have reshaped the input layer of the VGG-16 architecture to a shape of [60,60,3] and the model receives input as 60 $\times$ 60 pixel event images with three channels to leverage pre-trained ImageNet weights. There are five pooling layers in the model as per the architecture or five blocks with one pooling layer. The first two convolutional blocks have two convolutional layers, and the later ones have three convolutional layers each. Early layers deal with large images, so fewer layers are used to reduce computation. As the image size shrinks, more layers are added to capture complex features. This design builds depth gradually while keeping the model efficient and powerful. For the fully connected CNN model, we have utilized a single hidden layer architecture, created with Keras, a high-level deep learning library in Python \cite{keras}. Keras acts as a user-friendly interface for TensorFlow which is a library that facilitates efficient numerical computation and automatic differentiation. The output layer consists of five nodes, representing event classification with integer labels 0, 1, 2, 3 and 4, representing the direct, sequential decay branches and $\alpha$ + $^{40}$Ar, $\alpha$ + $^{12}$C and $\alpha$ + $^{16}$O scattering events, respectively, as shown in figure \ref{fig:m_vgg}.

%The efficiency of the track classification model using the VGG16 based CNN architecture has been compared for different pixel sizes to optimize the detector readout geometry {\bf(Did you discuss this in the paper or any result shown??)}

The original dataset, comprising of each class of nuclear events containing 1000 images, has been divided into training, validation and testing sets.
In the training stage, 70$\%$ of the dataset has been used for iterative optimization of the hyperparameters of the CNN model. The Adam (ADAptive Moment estimation) optimizer helps a CNN model learn by adjusting weights to reduce errors during training. It combines the benefits of momentum and adaptive learning rates, making learning faster and more stable. Adam automatically tunes the step size for each parameter, helping the model converge quickly and efficiently. While training the model, weights pre-trained on the ImageNet dataset were used to leverage learned features. The optimized hyperparameters are detailed in Table \ref{tab:parameters}.
To reduce the risk of overfitting, a regularization method has been implemented during the training phase, as mentioned before. 
\begin{table}[h!]
    \centering
    \begin{tabular}{|c|c|}
        \hline
            Epochs & 20 \\
        \hline
        Batch size & 128 \\
        \hline
        Learning rate & 0.001 \\
        \hline
        Loss & categorical crossentropy \\
        \hline
        Optimizer & Adam \\ 
        \hline
    \end{tabular}
    \caption{CNN model parameters and values}
    \label{tab:parameters}
    \end{table}
Then, 10$\%$ of the dataset has been used to validate the performance of the model after each 20 epochs. The remaining 20$\%$ of the original dataset has been used to estimate the performance of the trained CNN model by testing the predicted labels with actual labels.  

%\subsection{Model fitting methodology}
% \begin{figure}[h!]
 %\centering
 %    \includegraphics[scale=0.5]{Myvgg16.png}
 %\caption{CNN based on Vgg16 used for event classification}
  %   \label{m_vgg}
   %\end{figure} 
%VGG is a widely popular and highly efficient CNN architecture for image recognition which is very competitive even today compared to Res-Net or Dense-Net. 

\section{Performance Analysis}
\label{PerAnal}
In this section, the evaluation of the training and validation, followed by classification performance of the model, are discussed. The effect of variation in readout segmentation of the SAT-TPC has also been studied.

\subsection{Training \& Validation Accuracy}
\begin{figure}[h!]
     \centering
     \begin{subfigure}[b]{0.45\linewidth}
         \centering
         \includegraphics[width=\textwidth]{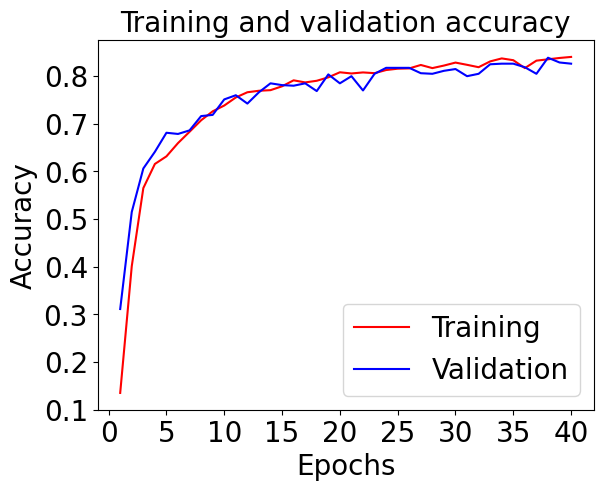}
         \caption{}
         \label{fig:drift}
     \end{subfigure}
     % \hspace{30 mm}
     \begin{subfigure}[b]{0.45\linewidth}
         \centering
         \includegraphics[width=\textwidth]{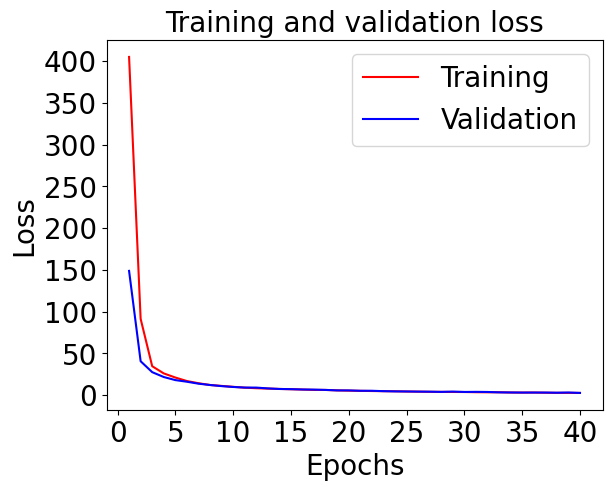}
         \caption{}
         \label{fig:dff}
     \end{subfigure}
   \caption{(a) Accuracy, and (b) loss of the training and validation processes}

        \label{l_curve}
\end{figure}

To evaluate the training of the present CNN model using a regularization technique, the accuracy and loss curves of training and validation have been produced, as shown in figure \ref{l_curve}. The curves show that the performance of the model has improved as the number of epochs has increased and stabilized after 10.

\subsection{Classification Performance}
We have evaluated the performance of the CNN model using three key metrics: precision, recall, and F1 score. These metrics have provided more detailed insights into its performance and errors than accuracy alone. For the track classification, we have focused on the ability of the model to identify direct and sequential decay events and the backgrounds accurately.
The precision is a measure of the accuracy of a model in identifying the positive instances which is calculated as the ratio of correctly predicted positive observations to the total predicted positives. In simpler terms, it tells us how many of the items that the model identifies as positive are actually positive. Recall, also known as sensitivity, measures the ability of the model to identify all relevant instances within a dataset. It is calculated as the ratio of correctly predicted positive observations to all observations in the actual class. Essentially, it indicates how many of the actual positives the model is able to capture. The F1 score is the harmonic mean of precision and recall. It provides a single metric that balances both the precision and recall of a model, especially useful when the classes are imbalanced. The F1 score combines the strengths of precision and recall, making it a comprehensive metric to evaluate the performance of a model. It is useful particularly in situations where the cost of false positives and false negatives is significant. The support column indicates the number of true instances (actual samples) of each event class in the dataset used to evaluate the model. The performance metrics of the present model are given in table \ref{tab:parametersp_matrice}. These have been determined with a readout segmentation of 2 mm $\times$ 2 mm.
\begin{table}[ht]
\centering
\caption{Classification report of VGG-16 based CNN model}
\label{tab:exp_exp_results}
\begin{tabular}{|l|l|c|c|c|c|c|}
\hline
\textbf{Readout segmentation} & \textbf{Event classification task} & \textbf{Precision} & \textbf{Recall} & \textbf{F1} & \textbf{Support} \\ \hline

 \multirow{5}{*}{2 mm $\times$ 2 mm} & direct decay & 0.70 & 0.65 & 0.67   & 200     \\ 
         \cline{2-6}

    & sequential decay & 0.68 & 0.72 & 0.70  & 200   \\
             \cline{2-6}
        & $\alpha$+$^{40}$Ar scattering & 0.98 & 0.98 & 0.98  & 200  \\
                     \cline{2-6}
        & $\alpha$+$^{12}$C scattering & 0.94 & 0.94 & 0.94  &  200 \\
                             \cline{2-6}
        & $\alpha$+$^{16}$O scattering & 0.92 & 0.94 & 0.93  & 200  \\
\hline    
\end{tabular}
    \label{tab:parametersp_matrice}

\end{table}

\begin{figure}[htb]
 \centering
\includegraphics[scale=0.3]{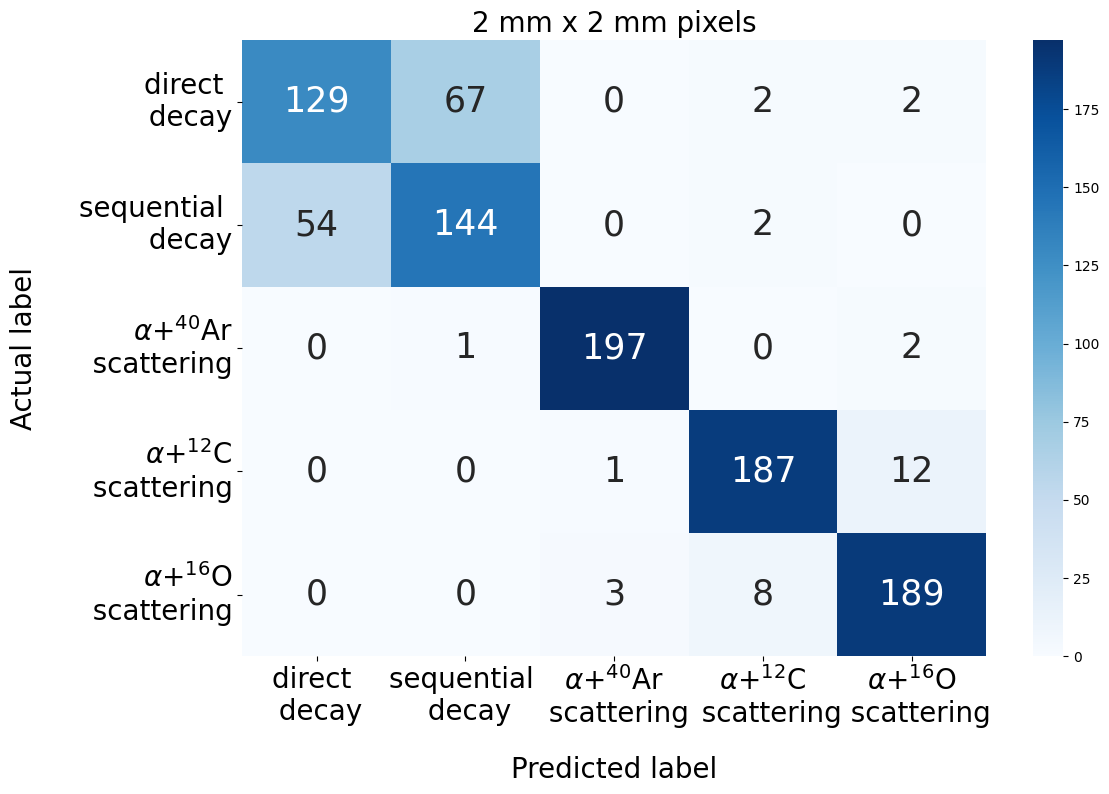}
 \caption{Confusion matrix}
 
     \label{cm}
   \end{figure}

The two metrics, precision and recall, are best understood by producing a confusion matrix, as shown in figure \ref{cm} which facilitates assessment of the performance and errors of the CNN model. It is evident from the matrix that the model has exhibited strong performance in classifying the scattering events $\alpha$ + $^{40}$Ar, $\alpha$ + $^{12}$C, and $\alpha$ + $^{16}$O, with high accuracy. The matrix reveals the misclassifications between the direct and sequential decays which indicates that these classes might require more distinct training data. In order to test this concept, the readout segmentation has been varied to study its effect on the performance of the model (discussed in the next sub-section). 

\subsection{Effect of Readout Segmentation} 
The scheme of the anode plane segmentation of the SAT-TPC in the simulation model has been varied to 1 mm $\times$ 1 mm and 3 mm $\times$ 3 mm pixels. The calculation has been carried out with the same class of events discussed in the previous section. The classification performance for the two segmentation schemes in terms of the three metrics is given in table \ref{pm_comparision}. It shows that the reduction in the size of the readout pixel from 3 mm $\times$ 3 mm to 1 mm $\times$ 1 mm has led to improved classification accuracy, particularly for decay events. This suggests that finer segmentation enhances spatial resolution, allowing the CNN to distinguish subtle differences in track morphology. 
\begin{table}[ht]
\centering
\caption{Classification report  for segmentation schemes 3 mm $\times$ 3 mm and 1 mm $\times$ 1 mm}
\label{tab:exp_exp_results}
\begin{tabular}{|l|l|c|c|c|c|c|}
\hline
\textbf{readout segmentation} & \textbf{event classification task} & \textbf{Precision} & \textbf{Recall} & \textbf{F1} & \textbf{Support} \\ \hline

 \multirow{5}{*}{3 mm $\times$ 3 mm} & direct decay & 0.62 & 0.67 & 0.64   & 200     \\ 
         \cline{2-6}

    & sequential decay & 0.64 & 0.57 & 0.61  & 200   \\
             \cline{2-6}
        &  $\alpha$+$^{40}$Ar scattering & 0.94 & 0.90 & 0.92  & 200  \\
                     \cline{2-6}
        & $\alpha$+$^{12}$C scattering & 0.86 & 0.85 & 0.86  &  200 \\
                             \cline{2-6}
        & $\alpha$+$^{16}$O scattering & 0.80 & 0.86 & 0.83  & 200  \\
\hline    
 \multirow{5}{*}{1 mm $\times$ 1 mm} & direct decay & 0.90 & 0.82 & 0.86   & 200     \\ 
         \cline{2-6}

    & sequential decay & 0.83 & 0.90 & 0.87  & 200   \\
             \cline{2-6}
        &  $\alpha$+$^{40}$Ar scattering & 0.99 & 1.00 & 1.00  & 200  \\
                     \cline{2-6}
        & $\alpha$+$^{12}$C scattering & 0.99 & 1.00 & 0.99  &  200 \\
                             \cline{2-6}
        & $\alpha$+$^{16}$O scattering & 0.99 & 0.98 & 0.99  & 200  \\
\hline    
\end{tabular}
    \label{pm_comparision}

\end{table}

\begin{figure}[htb]
    \centering
    \begin{subfigure}{0.48\textwidth}
        \centering
        \includegraphics[width=\textwidth]{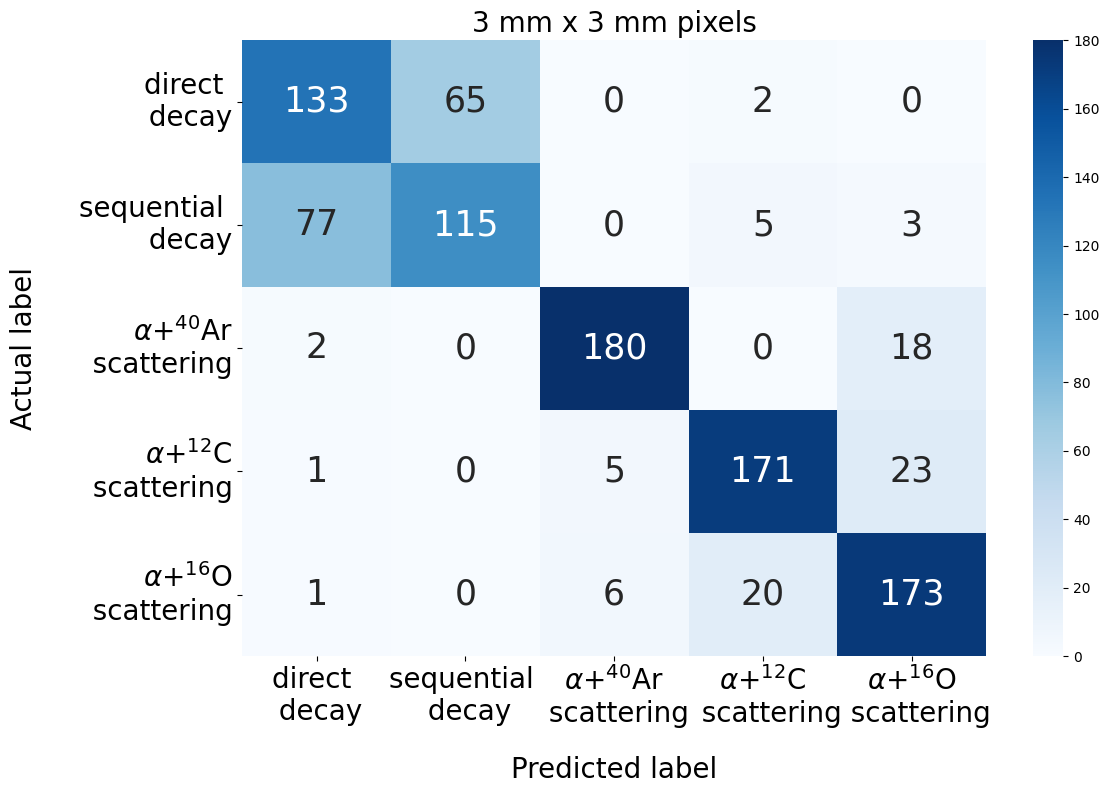}
        \caption{3 mm x 3 mm pixels}
        \label{fig:cm_3_3}
    \end{subfigure}
    \hfill
    \begin{subfigure}{0.48\textwidth}
        \centering
        \includegraphics[width=\textwidth]{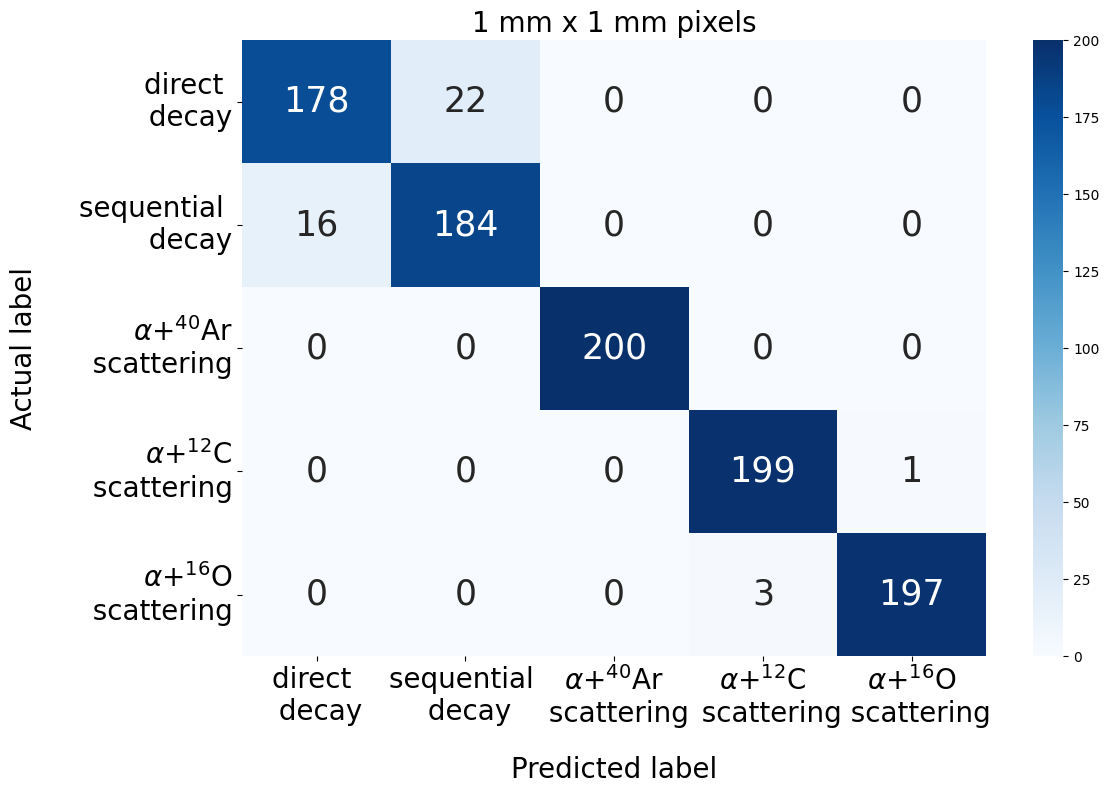}
        \caption{1 mm x 1 mm pixels}
        \label{fig:cm_2_2}
    \end{subfigure}
    \caption{Comparison of confusion matrices for different pixel sizes}
    \label{grid_pixel}
\end{figure}

The corresponding confusion matrices, as determined for two different readout segmentation schemes, are shown in figure \ref{grid_pixel}.  The results clearly indicate that the readout segmentation 1 mm $\times$ 1 mm provides better resolution with respect to other choices. However, the improvement in the accuracy of event classification has been reflected mostly in the case of distinguishing direct and sequential decay events.

\section{Summary \& Conclusion}
\label{Con}
% A neural network describes a large class of computational models inspired by the processes by which biological neurons process information. It consists of interconnected layers of nodes, or perceptrons, each of which processes input data and passes the result to the next layer. Neural networks are designed to recognize patterns and make predictions based on input data.
The classification of Hoyle state decay events and background elastic scattering events forms the central theme of the present study, showcasing the application of a supervised machine learning technique, specifically a CNN, for accurate segregation of the relevant class of events. The work has highlighted the challenges of distinguishing between direct decay, sequential decay, and background scattering events $\alpha$ + $^{40}$Ar, $\alpha$ + $^{12}$C, and $\alpha$ + $^{16}$O in the context of using 30 MeV $\alpha$-beam and a prototype SAT-TPC, filled with active gas target Ar + CO$_2$. We have performed Monte Carlo simulations to study the nuclear interactions leading to formation of the Hoyle state, following excitation through inelastic scattering of $\alpha$ from the active target nucleus $^{12}$C and the relevant elastic scattering of $\alpha$ from the target nuclei $^{40}$Ar, $^{12}$C and $^{16}$O.  The decay kinematics have been analyzed in a boosted frame, allowing for the extraction of energy and angular distributions of the emitted $\alpha$-particles. The same has been done for the elastic scattering events to study the energy and angular distributions of the scattered products. The primary ionization by the beam and the scattering and decay products has been simulated in Geant4 using a low-energy physics list that has been used to reconstruct the event tracks in the SAT-TPC active volume, convoluted with a diffusion parameter, and projected on 2D readout with segmentation of 2 mm $\times$ 2 mm. Using the VGG-16 CNN architecture, the model has classified the events on the basis of the track morphology images. Dropout was employed as a regularization technique to prevent overfitting and enhance the robustness of the model, achieving high precision, recall, and F1 scores across all categories.

Despite strong classification performance, the study has identified specific trends in misclassifications, particularly among the decay events. The classification performance of the model for the decay events has been found to improve with the finer readout segmentation as it has reduced the misclassification rate of the decay events. These results have proved the efficiency of the model in distinguishing Hoyle state decays from the background events while highlighting areas for further optimization, such as refining feature representations and detector configurations to enhance separation between closely related scattering categories. 
In the future, the size of the dataset will be increased well above the current choice of 1000
events/class to enhance training diversity and classification performance, especially for the
more ambiguous decay modes.
Additionally, since the CNN is trained on Monte Carlo data, further investigation will be needed to evaluate its robustness and performance on real detector data, including domain adaptation or transfer learning approaches.

This work has demonstrated the utilization of a CNN in analyzing experimental data of low-energy nuclear physics experiments performed with the SAT-TPC, providing a framework for advancing robust event classification in nuclear and astrophysical research. Future work will explore advanced architectures, such as ResNet, and transformer-based models, to further enhance classification performance. In addition, real-time processing capabilities will be investigated to enable integration with experimental data acquisition systems.

\acknowledgments
The authors acknowledge the support provided by their institute. They are thankful to Dr. Tapan Kumar Rana and Dr. Santu Manna of the Variable Energy Cyclotron Center, Kolkata and Prof. Satyaki Bhattacharya of Saha Institute of Nuclear Physics, Kolkata for valuable discussions and suggestions.

% \paragraph{Note added.} This is also a good position for notes added
% after the paper has been written.

% Bibliography

%% [A] Recommended: using JHEP.bst file
%% \bibliographystyle{JHEP}
%% \bibliography{biblio.bib}

%% or
%% [B] Manual formatting (see below)
%% (i) We suggest to always provide author, title and journal data or doi:
%% in short all the informations that clearly identify a document.
%% (ii) please avoid comments such as "For a review'', "For some examples",
%% "and references therein" or move them in the text. In general, please leave only references in the bibliography and move all
%% accessory text in footnotes.
%% (iii) Also, please have only one work for each \bibitem.

\end{document}